\documentclass[aps,prd,preprint,draft,showpacs,eqsecnum]{revtex4}
\usepackage{graphicx,epsf,amssymb}
\newcommand{\be}{\begin{equation}} 
\newcommand{\ee}{\end{equation}}

\newcommand{\bea}{\begin{eqnarray}} 
\newcommand{\eea}{\end{eqnarray}} 
\newcommand{\Tr}{{\rm Tr}}

\newcommand{\nn}{\nonumber}
 
\newcommand{\NeqFour}{{\cal N} =4}
\newcommand{\F}{{\cal{F}}}

\newif\ifdraft
\drafttrue
\newif\ifpreprint
\preprinttrue

\def\sect#1{section~{\ref{#1}}}

\def\fig#1{fig.~{\ref{#1}}}

\def\Tr{\, {\rm Tr}}

\def\NeqFour{{\cal N}=4}
\def\NeqOne{{\cal N}=1}
\def\NeqZero{{\cal N}=0}
\def\ns{n_{\mskip-2mu s}}
\def\nf{n_{\mskip-2mu f}}
\def\Nc{N_{c}}

\def\sandp#1.#2.#3{%
\left\langle\smash{#1}{\vphantom1}^{-}\right|{#2}%
\left|\smash{#3}{\vphantom1}^{+}\right\rangle}
\def\sandpp#1.#2.#3{%
\left\langle\smash{#1}{\vphantom1}^{+}\right|{#2}%
\left|\smash{#3}{\vphantom1}^{+}\right\rangle}
\def\sandmm#1.#2.#3{%
\left\langle\smash{#1}{\vphantom1}^{-}\right|{#2}%
\left|\smash{#3}{\vphantom1}^{-}\right\rangle}
\def\spab#1.#2.#3{\sandmm#1.#2.#3}
\def\spba#1.#2.#3{\sandpp#1.#2.#3}
\def\spaa#1.#2.#3.#4{\sandmp#1.{#2#3}.#4}
\def\spbb#1.#2.#3.#4{\sandpm#1.{#2#3}.#4}
\def\spa#1.#2{\left\langle#1\,#2\right\rangle}
\def\spb#1.#2{\left[#1\,#2\right]}
\def\spash#1.#2{\vphantom{\hat K}\spa{\smash{#1}}.{\smash{#2}}}
\def\spbsh#1.#2{\vphantom{\hat K}\spb{\smash{#1}}.{\smash{#2}}}
\def\lor#1.#2{\left(#1\,#2\right)}
\def\sand#1.#2.#3{%
\left\langle\smash{#1}{\vphantom1}^{-}\right|{#2}%
\left|\smash{#3}{\vphantom1}^{-}\right\rangle}
\def\sandpp#1.#2.#3{%
\left\langle\smash{#1}{\vphantom1}^{+}\right|{#2}%
\left|\smash{#3}{\vphantom1}^{+}\right\rangle}
\def\sandpm#1.#2.#3{%
\left\langle\smash{#1}{\vphantom1}^{+}\right|{#2}%
\left|\smash{#3}{\vphantom1}^{-}\right\rangle}
\def\sandmp#1.#2.#3{%
\left\langle\smash{#1}{\vphantom1}^{-}\right|{#2}%
\left|\smash{#3}{\vphantom1}^{+}\right\rangle}

\def\tr{\mathop{\rm tr}\nolimits}

\def\MSbar{$\overline{\rm MS}$}

\newbox\SlashedBox 
\def\slashed#1{\setbox\SlashedBox=\hbox{#1}
\hbox to 0pt{\hbox to 1\wd\SlashedBox{\hfil/\hfil}\hss}#1}
\def\hboxtosizeof#1#2{\setbox\SlashedBox=\hbox{#1}
\hbox to 1\wd\SlashedBox{#2}}

\newbox\charbox
\newbox\slabox
\def\s#1{{      % Feynman slash
        \setbox\charbox=\hbox{$#1$}
        \setbox\slabox=\hbox{$/$}
        \dimen\charbox=\ht\slabox
        \advance\dimen\charbox by -\dp\slabox
        \advance\dimen\charbox by -\ht\charbox
        \advance\dimen\charbox by \dp\charbox
        \divide\dimen\charbox by 2
        \raise-\dimen\charbox\hbox to \wd\charbox{\hss/\hss}
        \llap{$#1$}
}}

\def\eqn#1{eq.~(\ref{#1})}
\def\Eqn#1{Equation~(\ref{#1})}
\def\eqns#1#2{eqs.~(\ref{#1}) and~(\ref{#2})}

\def\qb{{\bar q}}

\def\e{\epsilon}
\def\eps{\epsilon}
\def\F#1#2{\,{{\vphantom{F}}_{#1}F_{#2}}}
\def\Gr{{\rm Gr}}

\def\sign{{\mathop{\rm sign}\nolimits}}
\def\lr{\leftrightarrow}
\def\li#1{{\mathop{\rm Li}\nolimits}_#1}
\def\Li{\mathop{\rm Li}\nolimits}
\def\Ls{\mathop{\rm Ls}\nolimits}
\def\Split{\mathop{\rm Split}\nolimits}
\def\tree{{\rm tree}}
\def\oneloop{{1 \mbox{-} \rm loop}}

\def\cg{c_\Gamma}

\def\mod{\mathop{\rm mod}\nolimits}
\def\Kh{{\hat K}}

\def\sandp#1.#2.#3{%
\left\langle\smash{#1}{\vphantom1}^{+}\right|{#2}%
\left|\smash{#3}{\vphantom1}^{+}\right\rangle}
\def\ksl{\s{k}}

\def\kslh{\s{\hat k}}

\def\Res{\mathop{\rm Res}}
\def\tlambda{{\tilde\lambda}}

\def\V{V}
\def\F{F}

\def\Ll{\mathop{\rm L{}}\nolimits}

\def\Lz{\mathop{\hbox{\rm L}}\nolimits_0}

\def\Cuth{{\widehat {C}}}
\def\CuthRat{{\widehat {CR}}}
\def\Remaining{{\widehat {R}}}

\def\rat{{\rm rat}}
\def\Ph{{\hat P}}
\def\Pb{{\overline P}}
\def\Vertex{R}
\def\DiagrammaticRational{R^D}
\def\PureCut{C}
\def\Res{\mathop{\rm Res}}
\def\rsn{r_S}

\newbox\ourfigbox
\def\SizedFigureWithCaption#1#2#3{%
\setbox\ourfigbox
  \hbox{\hss\epsfxsize #1 \epsfbox{#2}\hss}
\hbox to \wd\ourfigbox{\vbox{\noindent\copy\ourfigbox\break
\vskip -6mm      \hbox to \wd\ourfigbox{\hss#3\hss}}}
}

\def\llongrightarrow{%
\relbar\mskip-0.5mu\joinrel\mskip-0.5mu\relbar
     \mskip-0.5mu\joinrel\longrightarrow}
\def\inlimit^#1{\buildrel#1\over\llongrightarrow}
\def\dash{\hbox{-\kern-.02em}}

\begin{document}
\hfuzz 10 pt

%@twocolumnfalse\endcsname 

\ifpreprint
\noindent
UCLA/05/TEP/18
\hfill SLAC--PUB--11315
\hfill Saclay/SPhT--T05/114
\hfill hep-ph/0507005
\fi

\title{Bootstrapping Multi-Parton Loop Amplitudes in QCD%
\footnote{Research supported in part by the US Department of 
 Energy under contracts DE--FG03--91ER40662 and DE--AC02--76SF00515}}

\author{Zvi Bern}
\affiliation{ Department of Physics and Astronomy, UCLA\\
\hbox{Los Angeles, CA 90095--1547, USA}
%\\{\tt bern@physics.ucla.edu}
}

\author{Lance J. Dixon} 
\affiliation{ Stanford Linear Accelerator Center \\ 
              Stanford University\\
             Stanford, CA 94309, USA
%\\{\tt lance@slac.stanford.edu}
}

\author{David A. Kosower} 
\affiliation{Service de Physique Th\'eorique\footnote{Laboratory 
   of the {\it Direction des Sciences de la Mati\`ere\/}
   of the {\it Commissariat \`a l'Energie Atomique\/} of France.}, 
   CEA--Saclay\\ 
          F--91191 Gif-sur-Yvette cedex, France
%\\{\tt David.Kosower@cea.fr}
}

\date{July 2005}

\begin{abstract}
We present a new method for computing complete one-loop amplitudes,
including their rational parts, in non-supersymmetric gauge theory.
This method merges the unitarity method with on-shell recursion
relations.  It systematizes a unitarity-factorization bootstrap
approach previously applied by the authors to the one-loop amplitudes
required for next-to-leading order QCD corrections to the processes
$e^+e^- \rightarrow Z,\gamma^* \rightarrow 4$ jets and $pp \rightarrow
W + 2$ jets.  We illustrate the method by reproducing the one-loop
color-ordered five-gluon helicity amplitudes in QCD that interfere
with the tree amplitude, namely $A_{5;1}(1^-,2^-,3^+,4^+,5^+)$ and
$A_{5;1}(1^-,2^+,3^-,4^+,5^+)$.  Then we describe the construction of
the six- and seven-gluon amplitudes with two adjacent
negative-helicity gluons, $A_{6;1}(1^-,2^-,3^+,4^+,5^+,6^+)$ and
$A_{7;1}(1^-,2^-,3^+,4^+,5^+,6^+,7^+)$, which uses the
previously-computed logarithmic parts of the amplitudes as input.  We
present a compact expression for the six-gluon amplitude.  No loop
integrals are required to obtain the rational parts.
\end{abstract}

\pacs{11.15.Bt, 11.25.Db, 11.25.Tq, 11.55.Bq, 12.38.Bx \hspace{1cm}}

\maketitle

%\vskip2.0pc]

%% Section I %%%%%%%%%%%%%%%%%%%%%%%%%%%%%%%%%%%%%%%%%%%%%%%%%%%%%%%%%

\renewcommand{\thefootnote}{\arabic{footnote}}
\setcounter{footnote}{0}

%%%%%%%%%%%%%%%%%%%%%%%%%%%%%%%%%%%%%%%%%%%%%%%%%

\section{Introduction}
\label{IntroSection}

The approaching dawn of the experimental program at CERN's Large
Hadron Collider calls for theoretical support in a number of areas.  A
key ingredient in the quest to find and understand the new physics at
the TeV scale will be our ability to deliver precise predictions for
a variety of observable processes.  Fulfilling this demand will depend
in turn on having versatile tools for calculating multi-particle, loop-level
scattering amplitudes in the component gauge theories of the Standard
Model.  Tree-level amplitudes provide a first but
insufficient step.  The size and scale-variation of the strong
coupling constant imply that even for a basic quantitative
understanding, one must also include the one-loop
amplitudes which enter into next-to-leading order corrections to cross
sections~\cite{GloverReview}.  An important class of
computations are of perturbative QCD and 
QCD-associated processes.  Extending the set of available processes to
$W+\hbox{\rm\ multi-jet}$ production, and beyond, will demand computations
of new one-loop amplitudes in perturbative QCD.

In this paper we will describe a new approach to computing complete
one-loop scattering amplitudes in non-supersymmetric theories such as
QCD.  This approach systematizes a unitarity-factorization bootstrap
approach applied by the authors to the computation of the
one-loop scattering amplitudes needed for $Z \rightarrow 4$ jets and
$p p \rightarrow W + 2$ jets at next-to-leading order in the QCD
coupling~\cite{ZFourPartons}.  As in that paper, the cut-containing
logarithmic and polylogarithmic terms are computed using the unitarity
method~\cite{Neq4Oneloop,Neq1Oneloop,Massive,%
UnitarityMachinery,BCFII,BBSTQCD} and four-dimensional tree-level
amplitudes as input.  The remaining rational-function pieces 
are computed via a factorization bootstrap, 
in the form of an on-shell recurrence
relation~\cite{BCFRecurrence,BCFW,OnShellRecurrenceI,Qpap}.  (In
ref.~\cite{ZFourPartons} the rational functions were constructed
as ans\"atze with the assistance of the factorization limits, and
verified by numerical comparison to a direct Feynman diagram
computation.)

The unitarity method has proven to be an effective means of
computing the logarithmic and polylogarithmic terms in gauge theory
amplitudes at one and two loops.  In massless supersymmetric theories
the complete one-loop amplitudes may be determined from the
four-dimensional cuts~\cite{Neq1Oneloop}.  This method has been
applied in a variety of amplitude calculations in
QCD~\cite{ZFourPartons,SelfDualYM,DDimUnitarity,%
OneLoopSplitUnitarity,TwoLoopSplit,BMST} and in supersymmetric gauge
theories~\cite{Neq4Oneloop,Neq1Oneloop,BRY,NeqFourSevenPoint,NeqFourNMHV}.
A recent improvement to the unitarity method~\cite{BCFII} uses {\it
complex\/} momenta within generalized
unitarity~\cite{ZFourPartons,TwoLoopSplit,NeqFourSevenPoint}, and
allows a simple determination of box integral coefficients.  (The name
`generalized unitarity', as applied to amplitudes for massive
particles, can be traced back to ref.~\cite{Eden}.)  The unitarity
method has spawned a number of related techniques, include the very
beautiful application of maximally-helicity-violating (MHV) vertices
to loop calculations~\cite{BST,BBSTQCD} and the
use~\cite{CachazoAnomaly,BCF7} of the holomorphic
anomaly~\cite{HolomorphicAnomaly} to evaluate the cuts.  The unitarity
method can also be used to determine complete amplitudes, including
all rational pieces~\cite{Massive,SelfDualYM,DDimUnitarity,BMST} by
applying full $D$-dimensional unitarity, where $D = 4-2\e$ is the
parameter of dimensional regularization~\cite{HV}.  This approach
requires the computation of tree amplitudes where at least two of the
momenta are in $D$ dimensions.  For one-loop amplitudes containing
only external gluons, these tree amplitudes can be interpreted as
four-dimensional amplitudes but with massive scalars.  Recent work has
used on-shell recursive techniques~\cite{BCFRecurrence,BCFW} to extend
the number of known massive-scalar
amplitudes~\cite{BadgerMassive}.  At present, the
$D$-dimensional unitarity approach has been applied to all $n$-gluon
amplitudes with $n=4$~\cite{BMST} and to special helicity
configurations with $n$ up to 6~\cite{SelfDualYM,BMST}.

The somewhat greater complexity of the $D$-dimensional cuts suggests
that it is worthwhile to explore other methods of obtaining the
rational terms.  We have additional information about these terms,
after all, beyond the knowledge that their $D$-dimensional cuts are
$D$-dimensional tree amplitudes.  Because we know {\it a priori\/} the
factorization properties of the complete one-loop
amplitude~\cite{Neq4Oneloop,BernChalmers}, we also know the
factorization properties of the pure rational terms.  It would be good
to bring this information to bear on the problem.  This idea was
behind the `bootstrap' approach used in ref.~\cite{ZFourPartons}.  The
idea was used to produce compact expressions for the
$Z\rightarrow q\qb g g$ amplitudes.  However, it was not presented in
a systematic form, and indeed, for sufficiently complicated amplitudes 
it can be difficult to find ans\"atze
with the proper factorization properties.  This shortcoming has
prevented wider application of these ideas.

Recent progress in calculations of gauge-theory amplitudes has led us
to re-examine the bootstrap approach. 
This progress has been stimulated by
Witten's proposal of a weak-weak duality between $\NeqFour$
supersymmetric gauge theory and the topological open-string $B$ model
in twistor space~\cite{WittenTopologicalString}.  (The roots of the
duality lie in Nair's description~\cite{Nair} of the simplest gauge
theory amplitudes.)  Witten also made the beautiful conjecture that
the amplitudes are supported on a set of algebraic curves in twistor
space.  The underlying twistor structure of gauge theories, as revealed
by further
investigation~\cite{RSV,Gukov,BBK,CSWII,CSWIII,BBKR,CachazoAnomaly},
has turned out to be even simpler than originally conjectured.  (For a
recent review, see ref.~\cite{CSReview}.)  The underlying twistor
structure was made manifest by Cachazo, Svr\v{c}ek and Witten~\cite{CSW}, 
in a new set of diagrammatic rules for computing all tree-level
amplitudes, which use MHV amplitudes as vertices.
These MHV rules led to further progress in the computation of
tree-level~\cite{CSW,Higgs,Currents,RSVNewTree,
BCFRecurrence,BCFW,TreeRecurResults,BadgerMassive} amplitudes.
Brandhuber, Spence, and Travaglini~\cite{BST} provided
the link between
loop computations using MHV vertices and those done in the 
unitarity-based method.  This development in turn opened the way for further
computations and insight 
at one loop~\cite{BCF7,NeqFourSevenPoint,BCFII,NeqFourNMHV,
OtherGaugeCalcs,BBSTQCD}.  The remarkable conclusion of all
these studies is that gauge theory amplitudes, especially in
supersymmetric theories, are {\it much\/} simpler than had been
anticipated, even in light of known, simple, results.  Several
groups have also studied multi-loop amplitudes, and have
found evidence for remarkable simplicity, at least for maximal
supersymmetry~\cite{BRY}.

Recently, Britto, Cachazo and Feng wrote down~\cite{BCFRecurrence}
a new set of tree-level
recursion relations.  Recursion relations have long been
used in QCD~\cite{BGRecurrence,DAKRecurrence}, 
and are an elegant and efficient
means for computing tree-level amplitudes. The new recursion
relations differ in that they employ only {\it on-shell\/}
amplitudes (at {\it complex\/} values of the external momenta).  These
relations were stimulated by the compact forms of seven- and
higher-point tree
amplitudes~\cite{NeqFourSevenPoint,NeqFourNMHV,RSVNewTree} that
emerged from studying infrared consistency equations~\cite{UniversalIR}
for one-loop amplitudes.  
A simple and elegant
proof of the relation using special complex continuations
of the external momenta has been given by Britto, Cachazo, Feng and
Witten~\cite{BCFW}. 
Its application yields compact expressions for tree amplitudes
in gravity as well as gauge theory~\cite{TreeRecurResults}, and extends 
to massive theories as well~\cite{BadgerMassive}.  

In principle, recursion relations of this type could provide a
systematic way to carry out the factorization bootstrap at one loop.
One must however confront a number of subtleties in attempting to
extend them from tree to loop level.  The most obvious problem is that
the proof of the tree-level recursion relations relies on the
amplitudes having only simple poles; loop amplitudes in general have
branch cuts.  Moreover, the factorization properties of loop
amplitudes evaluated at complex momenta are not fully understood;
unlike the case of real momenta, there are no theorems specifying these
properties.  Indeed, there are double pole and `unreal'
pole contributions that must be taken into
account~\cite{OnShellRecurrenceI,Qpap}.

In a pair of previous papers~\cite{OnShellRecurrenceI,Qpap} we have applied
on-shell recursion relations to the study of finite one-loop
amplitudes in QCD.  These helicity amplitudes vanish at
tree level. Accordingly, the one-loop amplitudes are finite, and
possessing no four-dimensional cuts, are purely rational functions.
Through careful choices of shift variables and studies of known
amplitudes, we found appropriate double and unreal pole contributions 
for the recursion relations, and used them to recompute known gluon
amplitudes, and to compute fermionic ones for the first time.

While we will not give a derivation of complex factorization in 
the present paper, it is heartening that no new subtleties of this sort
arise in the amplitudes studied here, beyond those studied
in refs.~\cite{OnShellRecurrenceI,Qpap}. 
The systematization we shall present suggests that a proper and general
derivation of the complex factorization behavior should indeed be possible.

In this paper, we focus on the issue of setting up on-shell recursion
relations in the presence of branch cuts.  We describe a new method
for merging the unitarity technique with the on-shell recursion procedure.  
As mentioned above, we follow the procedure introduced in
ref.~\cite{ZFourPartons}, determining the cut-containing logarithms
and polylogarithms via the unitarity method, and then
determining the rational functions via a factorization bootstrap. We
derive on-shell recursion relations for accomplishing the bootstrap.
In general, both the rational functions and cut pieces have spurious
singularities which cancel against each other.  These spurious
singularities would interfere with the recursion because their
factorization properties are not universal.  We solve this problem by
using functions which are manifestly free of the spurious
singularities, at the price of adding some rational functions to the cut
parts.  These added rational functions have an overlap with the
on-shell recursion.  To handle this situation, we derive a recursion relation
which accounts for these overlap terms.

To illustrate our bootstrap method we recompute the rational-function
parts of the known~\cite{GGGGG} five-gluon amplitudes.
We present all the intermediate steps determining the rational 
functions of one of the five-gluon amplitudes, in order to 
underline the algebraic simplicity of the procedure.
As a demonstration of its utility, we also compute two new results,
the six- and seven-gluon amplitudes 
with two color-adjacent negative helicities.
We present the complete six-gluon amplitude in a compact form.
These results have all the required factorization properties in
real momenta, a highly non-trivial consistency check.  A computation
based purely on the unitarity method, that is to say based on 
full $D$-dimensional unitarity, would provide a further check.

This paper is organized as follows.  In the next section,
we review our notation and the elements entering into a decomposition
of QCD amplitudes at tree level and one loop.
In \sect{DerivationSection}, we derive a new
on-shell recursion-based formula for general one-loop amplitudes.
In \sect{ReviewSection}, we review the relevant known amplitudes,
and pieces thereof, and lay out the
vertices that will be used for the recomputation of the five-point
amplitude and the computation of the six- and seven-point amplitudes.
In \sect{FivePointSection}, we display the recomputation of the
five-point amplitude in great detail.  In \sect{SixPointSection},
we compute and quote the six-point amplitude, and present the
diagrams for the seven-point amplitude.  
We then give our conclusions.

%%%%%%%%%%%%%%%%%%%%%%%%%%%%%%%%%%%%%%%%%%%%%%%%%%%%%%%%

\vskip 15pt
\section{Notation}
\label{NotationSection}

In this section we summarize the notation used in the remainder of
the paper, following the notation of our previous
papers~\cite{OnShellRecurrenceI,Qpap}. We
use the spinor helicity formalism~\cite{SpinorHelicity,TreeReview}, 
in which the amplitudes are expressed in terms of spinor inner-products,
\begin{equation}
\spa{j}.{l} = \langle j^- | l^+ \rangle = \bar{u}_-(k_j) u_+(k_l)\,, 
\hskip 2 cm
\spb{j}.{l} = \langle j^+ | l^- \rangle = \bar{u}_+(k_j) u_-(k_l)\, ,
\label{spinorproddef}
\end{equation}
where $u_\pm(k)$ is a massless Weyl spinor with momentum $k$ and positive
or negative chirality.  We follow the convention that all legs are
outgoing. The notation used here follows the standard QCD
literature, with $\spb{i}.{j} = \sign(k_i^0 k_j^0)\spa{j}.{i}^*$ so
that,
\begin{equation}
\spa{i}.{j} \spb{j}.{i} = 2 k_i \cdot k_j = s_{ij}\,.
\end{equation}
These spinors are connected to Penrose's
twistors~\cite{Penrose} via a Fourier
transform of half the variables, {\it e.g.} the 
$u_-$ spinors~\cite{Penrose,WittenTopologicalString}.  
(Note that the QCD-literature square bracket $\spb{i}.{j}$ employed
here differs by an overall
sign compared to the notation commonly found in twistor-space
studies~\cite{WittenTopologicalString}.)  We also define,
as in the twistor-string literature,
\begin{equation}
 \lambda_i \equiv u_+(k_i), \qquad \tlambda_i \equiv u_-(k_i) \,.
\label{lambdadef}
\end{equation}

\def\vmu{{\vphantom{\mu}}}
We denote the sums of cyclicly-consecutive external momenta by
\begin{equation}
K^\mu_{i\cdots j} \equiv 
   k_i^\mu + k_{i+1}^\mu + \cdots + k_{j-1}^\mu + k_j^\mu \,,
\label{KDef}
\end{equation}
where all indices are mod $n$ for an $n$-gluon amplitude.
The invariant mass of this vector is $s_{i\cdots j} = K_{i\cdots j}^2$.
Special cases include the two- and three-particle invariant masses, 
which are denoted by
\begin{equation}
s_{ij} \equiv K_{i,j}^2
\equiv (k_i+k_j)^2 = 2k_i\cdot k_j,
\qquad \quad
s_{ijk} \equiv (k_i+k_j+k_k)^2 \,.
\label{TwoThreeMassInvariants}
\end{equation}
We also define spinor strings,
\begin{eqnarray}
\spab{i}.{(a+b)}.{j} &=& \spa{i}.{a} \spb{a}.{j} + \spa{i}.{b} \spb{b}.{j} \,,
          \nonumber   \\
\spbb{i}.{(a+b)}.{(c+d)}.{j} &=& 
     \spb{i}.{a} \spab{a}.{(c+d)}.{j} +
     \spb{i}.{b} \spab{b}.{(c+d)}.{j} \,,
\end{eqnarray}
and gamma matrix traces, 
\begin{eqnarray}
\tr_+[a\, b \, c \, d] & = & 
        \spb{a}.{b} \spa{b}.{c} \spb{c}.{d} \spa{d}.{a} \,, \\
\tr_+[a\, b \, c \, (d+e)] & = & 
        \spb{a}.{b} \spa{b}.{c} \spb{c}.{d} \spa{d}.{a} 
       + \spb{a}.{b} \spa{b}.{c} \spb{c}.{e} \spa{e}.{a} \,.
\end{eqnarray}

We use the trace-based color
decomposition of amplitudes~\cite{TreeColor,BGSix,MPX,TreeReview}.
For tree-level amplitudes with $n$ external gluons, this decomposition
is,
\begin{equation}
{\cal A}_n^\tree(\{k_i,h_i,a_i\}) = g^{n-2}
\sum_{\sigma \in S_n/Z_n} \Tr(T^{a_{\sigma(1)}}\cdots T^{a_{\sigma(n)}})\,
A_n^\tree(\sigma(1^{h_1},\ldots,n^{h_n}))\,,
\label{TreeColorDecomposition}
\end{equation}
where $g$ is the QCD coupling,
$S_n/Z_n$ is the group of non-cyclic permutations on $n$
symbols, and $j^{h_j}$ denotes the $j$-th gluon, with momentum 
$k_j$, helicity $h_j$, and adjoint color index $a_j$.
The $T^a$ are SU$(N_c)$ color matrices in the fundamental representation,
normalized so that $\Tr(T^a T^b) = \delta^{ab}$.  The color-ordered
amplitude $A_n^\tree$ is invariant under a cyclic permutation of its
arguments.

When all internal particles
transform in the adjoint representation of SU$(N_c)$, the color
decomposition for one-loop $n$-gluon amplitudes is given by~\cite{BKColor},
\begin{equation}
{\cal A}_n^{\rm adjoint} ( \{k_i,h_i,a_i\} ) = g^n
 \sum_{J} n_J  \sum_{c=1}^{\lfloor{n/2}\rfloor+1}
      \sum_{\sigma \in S_n/S_{n;c}}
     \Gr_{n;c}( \sigma ) \,A_{n;c}^{[J]}(\sigma) \,,
\label{AdjointColorDecomposition}
\end{equation}
where ${\lfloor{x}\rfloor}$ is the largest integer less than or equal
to $x$.  The sum $J\in \{0,1/2,1\}$ runs over all spins of particles
and $n_J$ is the multiplicity of each spin.  We assume all particles
are massless.  The leading-color structure,
\begin{equation}
\Gr_{n;1}(1) = N_c\ \Tr (T^{a_1}\cdots T^{a_n} ) \,, 
\end{equation}
is $N_c$ times the tree color structure.  The subleading-color
structures are given by
\begin{equation}
\Gr_{n;c}(1) = \Tr( T^{a_1}\cdots T^{a_{c-1}} )\,
\Tr ( T^{a_c}\cdots T^{a_n}) \,.
\end{equation}
In \eqn{AdjointColorDecomposition},
$S_n$ is the set of all permutations of $n$ objects, and $S_{n;c}$ is
the subset leaving $\Gr_{n;c}$ invariant.  For adjoint particles
circulating in the loop, the subleading-color partial amplitudes, 
$A_{n;c}$ for $c>1$, are
given by a sum over permutations of the leading-color
ones~\cite{Neq4Oneloop}.  Therefore we need to compute directly only
the leading-color, single-trace, partial amplitudes
$A_{n;1}(1^{h_1},\ldots,n^{h_n})$.  The leading-color amplitude is
again invariant under cyclic permutation of its arguments.

In QCD, of course, there can be fundamental representation quarks present
in the loop.  In this case only the single-trace color structure
contributes, but it is smaller by a factor $\Nc$. In general, 
scalars or fermions in the $\Nc+ \overline\Nc$ representation give
a contribution,
\begin{equation}
{\cal A}_n^{\rm fund} ( \{k_i,h_i,a_i\} ) = g^n
 \sum_{J = 0,1/2} {n_J \over \Nc}
      \sum_{\sigma \in S_n/Z_n}
     \Gr_{n;1}( \sigma ) \,A_{n;1}^{[J]}(\sigma) \,,
\label{FundamentalColorDecomposition}
\end{equation}
to the one-loop amplitude.
We use a supersymmetric convention in which the number of states
for a single complex scalar (squark) is $4 \Nc$,
in order to match the number of states of a Dirac fermion (quark). 

Helicity amplitudes that do not vanish at 
tree level develop infrared and ultraviolet divergences
at one loop.  We regulate these dimensionally.
Following ref.~\cite{GGGGG}, for the divergent one-loop amplitudes we write, 
\begin{eqnarray}
A_{n;1}^{[0]}  &=& \cg\Bigl( \V_n^s A^\tree_n + i \F_n^s \Bigr) \,,
 \label{An0}\\
A_{n;1}^{[1/2]} &=& -\cg \Bigl( (\V_n^f+\V_n^s) A^\tree_n
                                              + i (\F_n^f+\F_n^s)\Bigr)\,,
   \label{An1/2} \\
A_{n;1}^{[1]} &=& \cg \Bigl((\V_n^g+4\V_n^f+\V_n^s) A^\tree_n
                                              + i (4\F_n^f+\F_n^s) \Bigr)\,,
\label{An1}
\end{eqnarray}
where 
\begin{equation}
\cg = {1\over(4\pi)^{2-\eps}}
  {\Gamma(1+\eps)\Gamma^2(1-\eps)\over\Gamma(1-2\eps)}\,.
\label{cgdefn}
\end{equation}
The $V_n^x$ parts contain the divergences,  while the $F_n^x$ are
finite. (Of course, there is some ambiguity in the separation between
divergent and finite terms.)
These pieces have
a natural interpretation in terms of supersymmetric and non-supersymmetric
parts~\cite{GGGGG}, 
\begin{eqnarray}
A^{\NeqFour}_{n;1} &=& \cg  A_n^\tree \V_n^g \,, 
\label{ANeq4} \\
A^{\NeqOne}_{n;1} &=& - \cg  \Bigl( A_n^\tree \V_n^f + i \F_n^f \Bigr) 
                                                           \,,  
\label{ANeq1} \\
A^{\NeqZero}_{n;1} &=& \cg  \Bigl( A_n^\tree \V_n^s + i \F_n^s \Bigr) 
                                                            \,. 
\label{ANeq0} 
\end{eqnarray}
Here $A^{\NeqFour}_{n;1}$ sums over the contributions of an  $\NeqFour$ 
multiplet consisting of one gluon, four Majorana fermions, and three
complex scalars, all in the adjoint representation.
The $\NeqOne$ amplitude contains the contributions
of an adjoint chiral multiplet, consisting of one complex 
scalar and one Weyl fermion.  The non-supersymmetric amplitudes, 
denoted by $\NeqZero$, are just the contributions of a complex scalar
in the loop, $A^{\NeqZero}_{n;1} = A^{[0]}_{n;1}$.

The utility of separating QCD amplitudes into supersymmetric and
non-supersymmetric pieces follows from their differing analytic
properties. The supersymmetric pieces can be constructed completely
from four-dimensional unitarity cuts~\cite{Neq4Oneloop,Neq1Oneloop} and
have no remaining rational contributions (in the limit $\e\to0$).
The examples discussed in this paper, 
$n$-gluon amplitudes with two negative-helicity 
gluons, have been known for quite some 
time~\cite{Neq4Oneloop,Neq1Oneloop}.   (Such amplitudes
are often referred to as `MHV' in the supersymmetric cases,
because amplitudes with fewer (zero or one) negative-helicity 
gluons vanish.)
The logarithmic and polylogarithmic terms in the non-supersymmetric 
(scalar) pieces can also be obtained from four-dimensional cuts, or
from MHV vertices.
These cut-containing terms are also known for all $n$-gluon amplitudes with 
two negative-helicity gluons.  They were computed first for the 
case where the two negative-helicity gluons are 
color-adjacent~\cite{Neq1Oneloop}, and more recently for
the general case~\cite{BBSTQCD}.

Here we focus on the unsolved problem of computing the
rational-function terms in the $\NeqZero$ contributions in 
six and higher-point amplitudes, given the knowledge of
the logarithmic and polylogarithmic terms.
Finding an effective computational approach to the rational
terms in $A^{\NeqZero}_{n;1}$
is tantamount to solving the problem in QCD.

The leading-color QCD amplitudes are expressible in terms of
the different components in eqs.~(\ref{ANeq4})--(\ref{ANeq0}) via
\begin{eqnarray}
A_{n;1}^{\rm QCD} &=& 
\cg \Biggl[(\V_n^g+4\V_n^f+\V_n^s) A^\tree_n + i (4\F_n^f+\F_n^s) \nn\\
&&  \hskip 3 cm 
  - {\nf \over \Nc} \Bigl( A_n^\tree (\V_n^s + \V_n^f) + 
                                   i (\F_n^s + \F_n^f) \Bigr) \Biggr] \,,
\label{AnQCD}
\end{eqnarray}
where $\nf$ is the number of active quark flavors in QCD.  We will
present the formul\ae\ for unrenormalized amplitudes.  
To carry out an \MSbar\ subtraction, one should subtract
from the leading-color partial amplitudes $A_{n;1}$ the quantity
\begin{equation}
  c_\Gamma \left[{(n-2)\over2}{1\over\e}\left({11\over3}
  - {2\over3} {\nf\over N_c} - {1\over3}{\ns\over N_c} \right) \right]
  A_n^\tree\,,
\label{MSbarsubtraction}
\end{equation}
where we also included a term proportional to 
the number of active fundamental representation scalars $\ns$, 
which vanishes in QCD.  

We will need to consider additional objects (parts of amplitudes), beyond
the $V^x$ and $F^x$ defined here,
in order to construct and apply appropriate on-shell recursion
relations for one-loop amplitudes.  The definition of these objects
and derivation of the relations is the subject of the next section.

%%%%%%%%%%%%%%%%%%%%%%%%%%%%%%%%%%%%%%%%%%%%%%%%%%%%%%%%

\section{On-Shell Recursion Relations for Loop Amplitudes}
\label{DerivationSection}

On-shell recursion relations provide an effective means for
obtaining remarkably compact forms for 
tree-level amplitudes~\cite{RSVNewTree,BCFRecurrence,BCFW},
and have led to a variety of new 
results~\cite{TreeRecurResults,BadgerMassive}.
In previous work~\cite{OnShellRecurrenceI,Qpap}, we have shown how
to use on-shell recursion relations to compute all the finite loop amplitudes 
of QCD.  These special helicity amplitudes vanish at tree level.
Hence the one-loop amplitudes are free of infrared and 
ultraviolet divergences, and they are `tree-like'
in that they contain no cuts (absorptive parts) in four dimensions.
The derivation of these loop recursion relations is similar in spirit
to the tree-level case, but it does require the treatment of 
factorizations which differ from the `ordinary'
factorization in {\it real\/} momenta.

In this paper, we will extend the analysis of 
refs.~\cite{OnShellRecurrenceI,Qpap} to cut-containing one-loop 
amplitudes (for which the corresponding tree-level amplitudes do 
not vanish), deriving new recursion relations for the rational 
functions appearing in such amplitudes.  
The new recursion relations allow us to systematize the 
factorization bootstrap approach of ref.~\cite{ZFourPartons}.  
We assume that the cut-containing terms have already been determined 
via the unitarity method or some other means.  

%%%%%%%%
\subsection{Analytic behavior of shifted loop amplitudes}
\label{AnalyticSubsection}

The starting point for our analysis, as for the finite loop
amplitudes, is to consider~\cite{BCFW}
a complex-valued shift of the momentum of a pair of external 
particles in an $n$-point amplitude,
$k_j \to \hat{k}_j(z)$, $k_l \to \hat{k}_l(z)$. 
This shift is best described in terms of the spinor 
variables $\lambda$ and $\tlambda$ defined in \eqn{lambdadef},
\begin{equation}
\tlambda_j \rightarrow \tlambda_j - z\tlambda_l \,, 
\hskip 2 cm 
\lambda_l \rightarrow \lambda_l + z\lambda_j \,.
\label{SpinorShift}
\end{equation}
This $(j,l)$ shift maintains overall momentum conservation, because
\begin{equation}
\ksl_j + \ksl_l = \lambda_j \tlambda_j + \lambda_l \tlambda_l
\ \ \to\ \ 
\kslh_j + \kslh_l = \lambda_j (\tlambda_j - z\tlambda_l)
   + (\lambda_l + z\lambda_j) \tlambda_l
= \ksl_j + \ksl_l \,,
\label{momcons}
\end{equation}
as well as the masslessness of the external momenta,
$\hat{k}_j^2 = \hat{k}_l^2 = 0$.   
Denote the original $n$-point amplitude by $A_n \equiv A_n(0)$, 
and the shifted one by $A_n(z)$.   We wish to determine
$A_n(0)$ by making use of the analytic properties of $A_n(z)$.

\begin{figure}[t]
\centerline{\epsfxsize 2.0 truein\epsfbox{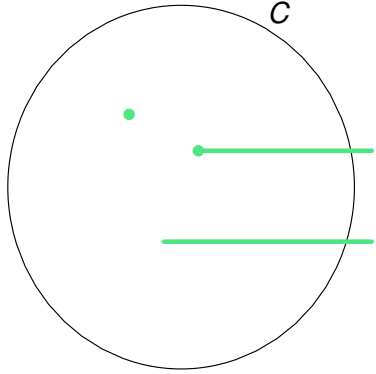}}
\caption{
A configuration of poles and branch cuts for a term in a one-loop 
amplitude.  The contour $C$ is a circle at $\infty$. }
\label{ExampleIntegralFigure}
\end{figure}

In the case of tree-level or finite one-loop amplitudes, $A_n(z)$ 
is a meromorphic function of $z$.  Here we also encounter
branch cuts, which may terminate at poles,
as depicted in \fig{ExampleIntegralFigure}.
Branch cuts arise from logarithms or polylogarithms in the amplitudes. 
Consider, for example, the scalar contributions to the five-gluon amplitude 
with color-ordered helicity assignment $({-}{-}{+}{+}{+})$, 
recalled in \eqn{Fs5}.  It contains a logarithm,
$\ln( (-s_{23})/(-s_{51}) )$, multiplied by a rational coefficient.  
If we perform a shift~(\ref{SpinorShift}) with $(j,l)=(1,2)$,
the logarithm becomes
\begin{equation}
\ln \Biggl({-s_{23}-z \sand1.3.2 \over -s_{51} + z \sand1.5.2} \biggr)
= \ln \Biggl( { \spb2.3 ( \spa2.3 + z \spa1.3 )
              \over \spa1.5 ( \spb1.5 - z \spb2.5 ) } \Biggr)
 \,.
\end{equation}
This function has two branch cuts in $z$, one starting at 
\begin{equation}
z = {\spb1.5\over\spb2.5},
\end{equation}
the other starting at
\begin{equation}
z = -{\spa2.3\over\spa1.3}.
\end{equation}
Because of the form of the rational coefficient of $\ln((-s_{23})/(-s_{51}))$
in this case, neither branch cut starts at a pole.

%%%%%% FIGURE %%%%%%%%%%%
%%%%%%%%%%%%%%%%%

We assume that $j$ and $l$ can be chosen so that $A_n(z)\rightarrow 0$
as $z\rightarrow\infty$.  We consider the following quantity,
\begin{equation}
{1\over 2\pi i} \oint_C {dz\over z}\,A_n(z) \,,
\label{BasicContourIntegral}
\end{equation}
where the contour integral is taken around the circle at $\infty$.
A typical configuration for a term in a one-loop amplitude is shown in
\fig{ExampleIntegralFigure}.
Even though the contour crosses branch cuts, the integral still vanishes,
because $A_n(z)$ vanishes at infinity.  
Unlike the rational cases studied previously, however, this does
not mean that it is given simply by a sum of residues at its poles.  We
need to include those contributions, of course; but we also need to 
integrate around the branch cuts, with special handling for poles 
at the end of branch cuts.

%%%%% FIGURE %%%%%%%%%
\begin{figure}[t]
\centerline{\epsfxsize 2.0 truein\epsfbox{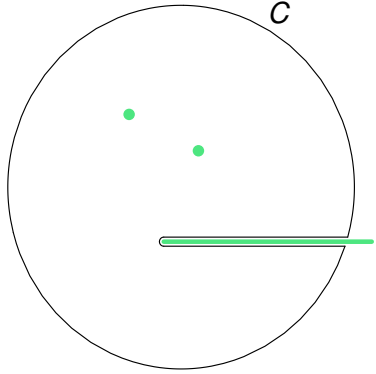}}
\caption{
A configuration of poles and branch cuts for a term in a one-loop amplitude,
with a branch-cut-hugging contour. 
}
\label{BranchCutIntegralFigure}
\end{figure}
%%%%%%%%%%%%%%%%%%%%%%%%%%%

\def\Disc{\mathop{\rm Disc}\nolimits}
We start with the `ordinary' branch cuts, with no pole touching
the branch cut.  We can imagine a
related contour, going along the circle at infinity, but avoiding the
branch cuts by integrating inwards along one side, and then outwards
along the other, as shown in \fig{BranchCutIntegralFigure}.  
(We will route the branch cuts so that no two overlap.)
The integral along {\it this\/} contour {\it is\/} given by the sum
of residues.  The difference between the two integrals is given by
the branch-cut-hugging integral,
\begin{equation}
{1\over2\pi i}\int_{B^\uparrow+i\epsilon} {dz\over z}\; A_n(z) 
+ {1\over2\pi i}\int_{B^\downarrow-i\epsilon} {dz\over z}\; A_n(z) \,,
\end{equation}
where $B^\uparrow$ is directed from an endpoint $B_0$ to infinity,
and $B^\downarrow$ is directed in the opposite way.  Now, $A_n(z)$ has
a branch cut along $B$, which means that it has a non-vanishing discontinuity,
\begin{equation}
2\pi i\Disc_B A_n(z) = A_n(z+i\epsilon)-A_n(z-i\epsilon),
\qquad z{\rm\ on\ }B \,.
\end{equation}

Thus our original vanishing integral can be written as follows,
\begin{equation}
0 = A_n(0) + \sum_{{\rm poles}\; \alpha} \Res_{z=z_\alpha} {A_n(z)\over z}
+ \int_{B_0}^\infty {dz\over z}\;\Disc_B A_n(z) \,.
\label{BasicAmplitudeEquation}
\end{equation}
The case with a pole at the end of the branch cut --- arising, for example from
terms containing $\ln(-s_{ab})/\spa{a}.b$ --- can be handled similarly, but
care must be taken with the evaluation of the integral along the branch cut.
From a conceptual point of view, we can also compute this case by
moving the pole away from the end of the branch cut by an amount $\delta$,
computing the branch-cut-hugging and residue terms separately, 
and taking the limit $\delta\rightarrow 0$ at the end.

%%%%%%%%
\subsection{Cut-containing terms, and their `completion'}
\label{CutContainingSubsection}

To proceed further, let us assume that we have already computed all terms
having branch cuts, plus certain closely related terms that can 
generally be obtained from the same computation.
That is, we have computed all polylog terms, 
all log terms, and all $\pi^2$ terms.  There are also
certain classes of rational terms that are natural to
include with the cut-containing terms.

In particular, there are rational terms whose presence is required to
cancel spurious singularities in the logarithmic terms.
Spurious singularities arise in the course of integral reductions.
They cannot be singularities of the final amplitude, because they 
are unphysical, and not singularities of any Feynman diagram.  
A simple example comes from a `two-mass' triangle integral for 
which two of the three external legs are off-shell (massive),
with momentum invariants $s_1$ and $s_2$, say.
When there are sufficiently many loop momenta 
inserted in the numerator of this integral, it gives rise to 
functions such as,
\begin{equation}
{\ln(r)\over (1-r)^2} \,,
\label{sampleL1pure}
\end{equation}
where $r$ is a ratio of momentum invariants (here $r=s_1/s_2$).
The limit $r\rightarrow 1$ (that is, $s_1\rightarrow s_2$) 
is a spurious singularity; it does not correspond to any physical
factorization. Indeed, this function always shows up in the amplitude
together with appropriate rational pieces,
\begin{equation}
{\ln(r)+1-r\over (1-r)^2} \,,
\label{sampleL1}
\end{equation}
in a combination which is finite as $r\rightarrow 1$.  
From a practical point of view, it is most convenient to `complete' 
the unitarity-derived answer for the cuts by replacing functions
like~\eqn{sampleL1pure} with non-singular combinations 
like~\eqn{sampleL1}.
Such completions are of course not unique; one could add 
additional rational terms free of spurious singularities.

There are other kinds of spurious singularities connected with
polylogarithms.  For example, in the scalar contributions to 
the five-gluon $(-{}+{}-{}+{}+)$ amplitude, there are factors of
$\spa2.4$ and $\spa2.5$ appearing in the denominators of certain
coefficients.  These might appear to give rise to non-adjacent collinear
singularities in complex momenta; but by expanding the polylogarithms
and logarithms in that limit, one can show that these singularities
are in fact absent.

Let us accordingly define two decompositions of the amplitude.  The first
is into `pure-cut' and `rational' pieces.
The rational parts are defined by setting all logarithms, 
polylogarithms, and $\pi^2$ terms to zero,
\begin{equation}
\Vertex_n(z) \equiv {1\over \cg} A_n\Bigr|_{\rm rat} = 
{1\over \cg} A_n\biggr|_{\ln, \Li, \pi^2 \rightarrow 0} \,.
\label{RationalDefinition}
\end{equation}
(Note that the normalization constant $\cg$, defined in \eqn{cgdefn},
plays no essential role in the following arguments, and is just
carried along for completeness.)
The `pure-cut' terms are the remaining terms, all of which must
contain logarithms, polylogarithms, or $\pi^2$ terms,
\begin{equation}
\PureCut_n(z) \equiv {1\over  \cg} A_n\Bigr|_{\rm pure-cut} = 
{1\over  \cg} A_n\biggr|_{\ln, \Li, \pi^2} \,.
\label{PureCutDefinition}
\end{equation}
In other words,
\begin{equation}
A_n(z) =  \cg \Bigl[ \PureCut_n(z) + \Vertex_n(z) \Bigr] \,,
\label{ACREq}
\end{equation}
where we have explicitly taken the ubiquitous one-loop factor 
$\cg$ outside of $\PureCut_n(z)$ and $\Vertex_n(z)$.

The second decomposition uses the 
`completed-cut' terms, obtained from $\PureCut_n(z)$
by replacing logarithms and polylogarithms by corresponding
functions free of spurious singularities.  We shall
call this completion $\Cuth_n$.  The decomposition defines the remaining
rational pieces $\Remaining_n$,
\begin{equation}
A_n(z) =  \cg \Bigl[ \Cuth_n(z) + \Remaining_n(z) \Bigr] \,.
\label{CompletedCutDecomposition}
\end{equation}
We also need to define the rational part of the completed-cut
terms, $\widehat{CR}_n(z)$.  We write,
\begin{equation}
\Cuth_n(z) = C_n(z) + \widehat{CR}_n(z) \,,
\label{CuthCCR}
\end{equation}
where
\begin{equation}
\CuthRat_n(z) \equiv \Cuth_n(z)\Bigr|_{\rm rat} \,.
\label{CuthRatDefinition}
\end{equation}
Combining eqs.~(\ref{ACREq}), (\ref{CompletedCutDecomposition}),
and (\ref{CuthCCR}), we see that the full rational part
is the sum of the rational part of the completed-cut terms,
and the remaining rational pieces,
\begin{equation}
\Vertex_n(z) = \widehat{CR}_n(z) + \Remaining_n(z)\,.
\label{RCRRhat}
\end{equation}

Now, because we know all the terms containing branch
cuts, we could compute the branch-cut-hugging integral,
\begin{equation}
\int_{B_0}^\infty {dz\over z}\;\Disc_B \Cuth_n(z) \,.
\end{equation}
However, there is no need to do the integral 
explicitly, because we already know the answer for the integral, plus the
corresponding residues.  It is just $\Cuth_n(0)$, part of the final
answer.  That is, applying the same logic to $\Cuth_n(z)$
as was applied to $A_n(z)$ in \eqn{BasicAmplitudeEquation}, we have,
\begin{equation}
 \Cuth_n(0) = 
- \sum_{{\rm poles}\; \alpha} \Res_{z=z_\alpha} {\Cuth_n(z)\over z}
- \int_{B_0}^\infty {dz\over z}\;\Disc_B \Cuth_n(z) \,.
\label{BasicCuthEquation} 
\end{equation}
In arriving at this result, we need to assume that $\Cuth_n(z)$,
in addition to the full amplitude, also vanishes as $z\rightarrow\infty$.  
(This constraint may place some restrictions on the allowable rational
completions of the cut terms.)

Using \eqn{BasicAmplitudeEquation}, 
the split-up~(\ref{CompletedCutDecomposition}),
and \eqn{BasicCuthEquation} to evaluate the terms involving 
$\Cuth_n(z)$, we can write our desired answer as follows,
\begin{eqnarray}
A_n(0) &=& 
- \cg \Biggl[ \int_{B_0}^\infty {dz\over z}\;\Disc_B \Cuth_n(z)
       +\sum_{{\rm poles}\;\alpha} \Res_{z=z_\alpha} {\Cuth_n(z)\over z}
       +\sum_{{\rm poles}\;\alpha} \Res_{z=z_\alpha} 
                 {\Remaining_n(z)\over z} \Biggr] 
\nonumber\\
&=&  \cg \Biggl[ \Cuth_n(0)
       -\sum_{{\rm poles}\; \alpha} \Res_{z=z_\alpha} 
                 {\Remaining_n(z)\over z}  \Biggr] 
\,.
\label{FormII}
\end{eqnarray}
Because we have completed the cut terms so that $\Cuth_n(z)$ contains 
no spurious singularities, the sums over the poles in \eqn{FormII}
are only over the genuine, `physical' poles in the amplitude.  
(As explained elsewhere, these are the poles that arise
for {\it complex\/} momenta, and not merely those that arise for {\it real\/}
momenta.)

%%%%
\subsection{Separate factorization of pure-cut and rational terms}
\label{SeparateFactorizationSubsection}

The residues of the completed-cut terms $\Cuth_n(z)$ at the 
genuine poles contain both rational and cut-containing functions.  
The residues of cut-containing functions necessarily have cut-containing
functions (and the residues of $\pi^2$-containing terms will necessarily
have factors of $\pi^2$, etc.), so that they will arise from cut-containing
parts of the factorized amplitudes.  The intuition from collinear
factorization of one-loop amplitudes suggests, however, that the 
pure-cut terms arise from pure-cut terms, and the rational terms 
arise from rational terms.  The purpose of this subsection is
to flesh out this intuition of separate factorization for
pure-cut and rational terms.

More concretely, the arguments of the logarithms or polylogarithms
are ratios of invariants $s_{i\cdots j}$.
(In a limited number of logarithms, the arguments are ratios of invariants 
to the renormalization scale squared.)  When we shift these ratios by shifting
momenta according to~\eqn{SpinorShift}, one of three things can happen to any
specific argument: 
\begin{enumerate}
\item The ratio may be invariant under the shift, so
that the residue simply has the original cut-containing function in it;
\item 
the ratio may acquire a dependence on $z$, but neither vanish nor
diverge at any of the poles in $z$ of the given term; 
\item 
or the ratio may
acquire a dependence on $z$, and either vanish or diverge at one of
the poles in the given term. 
\end{enumerate}
Because the branch cuts in a massless
theory start at either a vanishing
ratio or at a vanishing of its inverse, the last situation corresponds
to a pole touching the end of a branch cut.

In the first of these three cases, the cut-containing function clearly
arises from a cut-containing function in the residue of the pole; and
any associated rational terms arise from rational terms in the
residue.  In the second case, behavior under factorization in {\it
real\/} momenta suggests that we need consider only single poles.
In particular, for multiparticle invariants, even for complex momenta
we have only single poles to consider.  For two-particle channels,
double poles can arise (for complex
momenta) in the kinematic invariant of the momenta of two 
nearest-neighboring legs in the color ordering, but only at one loop
and only for certain helicity configurations.

In two-particle factorizations, one-loop splitting amplitudes
appear in addition to tree-level ones.  The remaining amplitude
left behind in the former case, however, is a tree-level amplitude,
which is of course purely rational.  In the latter case, we
need consider only single poles.  

We are left to consider the separation inside the 
cut-containing one-loop
splitting amplitudes themselves.
Because we are only interested in the scalar loop contributions, the
relevant splitting amplitudes we need are scalar-loop 
ones~\cite{Neq4Oneloop,Neq1Oneloop,BernChalmers,
OneLoopSplitUnitarity,BDSSplit}.
For the helicity configurations for which the tree-level splitting 
amplitude $\Split^{\tree}_{-\lambda}(x; a^{\lambda_a},b^{\lambda_b})$
is non-vanishing (where $\lambda$ is the helicity of the outgoing
off-shell leg, and $x$ is the longitudinal-momentum fraction carried
by leg $a$), we write the loop splitting amplitude as,
\begin{equation}
\Split^{\oneloop,\,[0]}_{-}(x; a^{\lambda_a},b^{\lambda_b}) =  
 \rsn^{\oneloop \, \lambda_a\lambda_b, \, [0]}(x, s_{ab})
 \times {\Split}^{\tree}_{-}(x; a^{\lambda_a},b^{\lambda_b}) \,,
\label{OnelooprS}
\end{equation}
where
\begin{eqnarray}
\rsn^{\oneloop \, -+, \, {[0]}}(x,s) & = & 0 \,,
\label{OneloopSplitExplicitmp}\\
\rsn^{\oneloop \, ++, \, {[0]}}(x,s)  &=& 
 \cg \, \biggl({\mu^2\over-s}\biggr)^{\eps} \,
    {2 x (1-x)\over(1-2\eps)(2-2\eps)(3-2\eps)}
     \,. ~~~~
\label{OneloopSplitExplicitpp} 
\end{eqnarray}
The vertex corresponding to the first splitting amplitude will
indeed vanish, and so it will not affect the separation between cut-containing
and rational pieces.  The situation with the second helicity configuration
remains to be studied.  Indeed, it appears in this case that for
complex momenta, there are genuinely non-factorizing contributions
in addition to pole contributions.  The shift choices we will make
in this paper avoid the appearance of this vertex, and so we postpone
the analysis of this case to future work.

Of course, outside of cut-containing terms,
double poles do arise in general, as
indicated by the finite one-loop splitting amplitude (the helicity
configuration that vanishes at tree level),
\begin{equation}
\Split^{\oneloop,\,[0]}_+(x; a^+,b^+) = -{1\over48\pi^2}
 \sqrt{x (1-x)} { \spb{a}.b \over \spa{a}.b^2} \,.  
\end{equation}
They would be handled as in
refs.~\cite{OnShellRecurrenceI,Qpap}; given the shift choices
we will make in the present paper, they will in any event not arise.

So long as we are indeed
considering only single poles, extracting the residue will not force
us to expand $\Cuth_n(z)$ in a series around the pole.
In this case, the two types of contributions, pure-cut and rational,
will remain separate in this class of contributions.  As explained
above, the third and last case, that of the pole hitting the branch
cut, can effectively be reduced to the second case by artificially
separating the pole from the end of the branch cut, and removing the
separation at the end of the calculation. 

As an example of the different kinds of behavior, consider the following
expression,
\begin{equation}
{\spa2.3\spb3.4^2 \spa4.1^2\spa1.2\spb1.5
 \over \spa3.4\spa4.5}
{\Ll_1\biggl( {-s_{23}\over -s_{51}} \biggr)\over s_{51}^2} \,,
\end{equation}
where $\Ll_1(r) = (\ln(r)+1-r)/(1-r)^2$,
under a $(1,5)$ and $(3,4)$ shift respectively.  Under the first
shift, we obtain a simple pole at $z = -\spa4.5/\spa4.1$, but the
argument of the $\Ll_1$ function is unchanged, and accordingly, the
logarithm and rational parts are left unaltered.  In the context
of a factorization, we would interpret the logarithm as arising from
a logarithm in the factorized amplitude, and the rational part inside
the $\Ll_1$ from a rational term.

Under the second shift, we obtain a simple pole at $z=-\spa4.5/\spa3.5$.
The argument of $\Ll_1$ at the pole is now,
\begin{equation}
{-\spa2.3(\spb3.2 - z_{\rm pole}\spb4.2)
\over -s_{51}} = 
{\spa2.3(\spa5.3\spb3.2 +\spa5.4 \spb4.2)
\over - \spa3.5  s_{51}}
= {\spa2.3 \spa1.5 \spb1.2
\over -\spa3.5 s_{51}} \,,
\end{equation}
but because the pole is a simple one, the separation between logarithm
and rational part is again left undisturbed.  If instead the original
expression had contained not a $\spa4.5$ in the denominator but rather
a $\spa4.5^2$, we would have had to expand the logarithm inside
$\Ll_1$ in order to extract the residue, and this expansion would have
led to rational terms whose origin was in a `pure-cut' term, rather
than a rational one, mixing up the two types of contributions. Fortunately, 
such combinations never arise in the analysis of the amplitudes discussed
in this paper.

\subsection{Residues of the remaining rational pieces $\Remaining_n(z)$}
\label{RemainingRatSubsection}

Thanks to the analysis of the previous subsection,
we can simplify our calculation by separating
the two classes of terms --- pure-cut and rational ---
in the factorized amplitudes.  Because we already know the 
cut-containing pieces, we need to analyze
only the rational terms.  (This separation also avoids the need to treat
poles hitting the ends of branch cuts explicitly.)  

We now examine more carefully the residues of the poles of 
the purely rational terms, $\Vertex_n(z)$,
and related to that, 
of the shifted remaining rational terms, $\Remaining_n(z)$,
appearing in \eqn{FormII}.
Given the $(j,l)$ shift~(\ref{SpinorShift}), we
define a partition $P$ to be a set of two or more cyclicly-consecutive
momentum labels containing $j$, such that the complementary set $\Pb$
consists of two or more cyclicly-consecutive labels containing $l$:
\begin{eqnarray}
 P &\equiv& \{ P_1, P_2, \ldots, j, \ldots, P_{-1} \} \,, 
\label{PartitionDef} \\
 \Pb &\equiv& \{ \Pb_1, \Pb_2, 
  \ldots, l, \ldots, \Pb_{-1} \} \,, \nn \\
 P &\cup& \Pb = \{ 1,2,\ldots,n \} \,. \nn
\end{eqnarray}
This definition ensures that the sum of momenta in each partition is 
$z$-dependent, so that it can go on shell for a suitable value of $z$. 
At tree level, the sum over residues becomes,
\begin{eqnarray}
 - \sum_{{\rm poles}\; \alpha} \Res_{z=z_\alpha} {A_n^\tree(z)\over z}
&=& 
A_n^\tree(k_1,\ldots,k_n)  \nonumber\\
&=& \sum_{{\rm partitions}\; P} \sum_{h = \pm}
A_L^\tree(k_{P_1},\ldots,\hat k_j,\ldots,k_{P_{-1}},-\Ph^h) \nonumber\\
&&\hphantom{\sum}\times
{i\over P^2-m_P^2} \times
A_R^\tree(k_{\Pb_1},\ldots,\hat k_l,\ldots,k_{\Pb_{-1}},\Ph^{-h})\,.
\label{GeneralRecursion}
\end{eqnarray}
The complex on-shell momenta $\hat k_j$,
$\hat k_l$ and $\Ph$ are determined by solving the on-shell condition,
$\Ph^2 = m_P^2$, for $z$.  Although the examples we discuss in this
paper are for massless particles, we allow for a mass $m_P^2$ to indicate 
that there is no special restriction to massless amplitudes, 
as has already been noted at tree level~\cite{BadgerMassive}.

At one loop, the sum analogous to \eqn{GeneralRecursion}
will have an additional two-fold sum.  In each term in this sum, 
either $A_L$ or $A_R$ will be a tree amplitude, and the other one will 
be a loop amplitude; in general both terms will appear.  
Taking the rational parts of the one-loop amplitudes appearing in this
expression, the one-loop physical-pole recursion for the 
rational terms is,
\begin{eqnarray}
 - \sum_{{\rm poles}\; \alpha} \Res_{z=z_\alpha} {\Vertex_n(z)\over z}
&\equiv& 
\DiagrammaticRational_n(k_1,\ldots,k_n) \nonumber\\
 &=& \hskip -.1cm
 \sum_{{\rm partitions}\, P}\, \sum_{h = \pm} \Biggl\{
\Vertex(k_{P_1},\ldots,\hat k_j,\ldots,k_{P_{-1}},-\Ph^h)  \nonumber\\
&& \null \hskip 2.5 cm \times
{i\over P^2} \times
A^\tree(k_{\Pb_1},\ldots,\hat k_l,\ldots,k_{\Pb_{-1}},\Ph^{-h}) \nn\\
&& \null \hskip 2.5 cm 
+ A^\tree(k_{P_1},\ldots,\hat k_j,\ldots,k_{P_{-1}},-\Ph^h)  
      \label{RationalRecursion} \\
&& \null \hskip 2.5 cm \times
{i\over P^2} \times
\Vertex(k_{\Pb_1},\ldots,\hat k_l,\ldots,k_{\Pb_{-1}},\Ph^{-h}) 
\Biggr\}
 \,, \hskip 1.3 cm  \nn
\end{eqnarray}
where we now assume that the intermediate states are massless.  This
result follows directly from the general factorization behavior of
one-loop amplitudes, plus the separate factorization of pure-cut and
rational terms that was established in the previous subsection.  
Just as in the case of the tree-level recursion (\ref{GeneralRecursion}),
it exhibits the required factorization properties in each channel $P$
(dropping the terms with logarithms, polylogarithms, and $\pi^2$).
Although the $R$ functions are not complete amplitudes, they can be
thought of as vertices from a diagrammatic perspective, and this
equation lends itself to the same kind of diagrammatic interpretation
available for \eqn{GeneralRecursion}.

However, the factorization cannot distinguish between the rational
terms we have already included in the completed-cut terms and the
remaining ones. That is, there would be an overlap or double count if we
were simply to combine the recursive diagrams with the completed-cut
terms.  To remove this overlap, we separate the physical-pole
contributions into those already included in the completed-cut terms
and those in the remaining rational terms.  Using \eqn{RCRRhat},
we know that
\begin{equation}
 - \sum_{{\rm poles}\; \alpha} \Res_{z=z_\alpha} 
          {\Vertex_n(z)\over z} 
= \DiagrammaticRational_n 
= 
       -\sum_{{\rm poles}\; \alpha} \Res_{z=z_\alpha} 
          {\CuthRat_n(z)\over z} 
%\nonumber\\ &\hphantom{=}&
       -\sum_{{\rm poles}\; \alpha} \Res_{z=z_\alpha} 
                 {\Remaining_n(z)\over z} \,.
\end{equation}
Because we know the completed-cut terms $\Cuth_n(z)$ and their rational parts
$\CuthRat_n(z)$ explicitly, we can compute the first term on
the right-hand side, and solve for the remaining terms using the
determination of $\DiagrammaticRational_n$ via
\eqn{RationalRecursion}.  Inserting the result into \eqn{FormII} then 
gives us the basic on-shell recursion relation for complete one-loop
amplitudes,
\begin{equation}
A_n(0) =  \cg \biggl[\Cuth_n(0)
+ \DiagrammaticRational_n 
 +\sum_{{\rm poles}\; \alpha} \Res_{z=z_\alpha} {\CuthRat_n(z)\over z}
\biggr]\,.
\label{BasicBootstrapEquation}
\end{equation}
To compute with this equation, we
construct $\DiagrammaticRational_n$ via `direct recursion' diagrams; that is,
via \eqn{RationalRecursion}.
We call the elements of the last term `overlap' terms.  Because
each pole is associated with a specific diagram, these can also
be given a diagrammatic interpretation.
Although the definition of the completed-cut terms $\Cuth_n$ is not
unique, the ambiguity cancels between $\Cuth_n(0)$ and the sum
over $\CuthRat_n$ residues.  In the calculations
in this paper, an astute choice of completion terms can simplify the
calculation, by simplifying the extraction of the residues in
the last term.

The reader may wonder how the calculation would have proceeded if we had
started with `pure' cut terms, not including any of the rational
pieces needed to eliminate the spurious singularities.  In this case,
the intermediate stages, and in particular \eqn{FormII}, would have included
a sum over the spurious singularities as well.  Since these
singularities include double and triple poles, we {\it would\/} have needed
to expand the logarithms in extracting the residues for the `overlap'
terms, and this expansion would have produced rational terms.  In contrast,
with our approach, we never evaluate residues at values of $z$ 
corresponding to unphysical spurious singularities, and we never have
to expand logarithmic functions.

In \sect{NotationSection}, we separated the one-loop amplitudes into
divergent and finite parts.  The amplitude as a whole satisfies the
bootstrap relation~\eqn{BasicBootstrapEquation}; but it turns out that
for the amplitudes we consider in the present paper, it can be applied
separately to the $V$ and $F$ terms.  As the recursion relation for
the pure-scalar parts of the former are basically the same as at tree
level, we will focus on the computation of the $F$ terms.  For this
purpose, we shall use quantities analogous to $\PureCut_n$,
$\Vertex_n$, $\Cuth_n$, $\Remaining_n$, and $\CuthRat_n$, defined as in
eqs.~(\ref{PureCutDefinition})--(\ref{CuthRatDefinition}), but with
respect to $F_n^s$ of \eqn{An0} instead of $A_n$.  Note that this shift 
of convention generates a relative factor of $i$ in the quantities we
use below, due to the relative $i$ in \eqn{An0}.

%%%%%%%%%%%%%%%%%%%%%%%%%%%%%%%%%%%%%%%%%%%%%%%%%%%%%%%%
\section{Review of Known Results}
\label{ReviewSection}

In this section we summarize the previously-computed amplitudes,
and pieces thereof, that feed into our recursive construction.
In this paper, we consider $n$-gluon amplitudes with two color-adjacent 
negative helicities, $A_{n;1}(1^-,2^-,3^+,4^+,\ldots,(n-1)^+,n^+)$.
As mentioned in \sect{NotationSection}, the $\NeqFour$
and $\NeqOne$ components of these amplitudes have been 
known for a while, so the only issue is the computation 
of the $\NeqZero$ or scalar contribution,
\begin{equation}
A_{n;1}^{[0]}(1^-,2^-,3^+,4^+,\ldots,(n-1)^+,n^+) \,.
\label{NearestNeighborn}
\end{equation}
Assigning intermediate helicities to all possible factorizations 
of this amplitude, as encountered in \eqn{RationalRecursion},
allows us to determine which lower-point amplitudes are required 
as input.  Besides the Parke-Taylor tree amplitudes with two 
(adjacent) negative helicities~\cite{ParkeTaylor,BGSix,MPX},
we shall need the one-loop scalar contributions with
one negative helicity~\cite{Mahlon}, for which we recently found
a compact form~\cite{Qpap}.  
The one-loop amplitudes with two adjacent negative helicities and
smaller values of $n$ are also needed, and are obtained recursively,
given a suitable starting point.

By choosing to shift the two negative-helicity legs we can avoid some
factorizations.  For example, for a generic choice of $j$ and $l$ in
the $(j,l)$ shift~(\ref{SpinorShift}), the amplitude could factorize
onto products containing a one-loop amplitude with all positive
external helicities and an internal helicity of either sign.  We avoid
these factorizations by choosing to shift the two negative-helicity
legs, $(j,l)=(1,2)$.
One external negative helicity then appears in each partition.
In addition to lower-point amplitudes, we also need the logarithmic
parts of the $n$-point amplitude~(\ref{NearestNeighborn}), obtained in
ref.~\cite{Neq1Oneloop}.  (The logarithmic parts of the more general 
set of amplitudes for two
non-adjacent negative helicities were recently obtained in
ref.~\cite{BBSTQCD}.)

In \sect{FivePointSection} we recompute the five-gluon helicity 
amplitude $A_{5;1}^{[0]}(1^-,2^-,3^+,4^+,5^+)$. 
For that computation, we take the four-point amplitude
$A_{4;1}^{[0]}(1^-,2^-,3^+,4^+)$ as an input,
but here we quote the previous five-gluon result too, in order to demonstrate
that the method works.  (We also outline the 
recursive construction of $A_{5;1}^{[0]}(1^-,2^+,3^-,4^+, 5^+)$,
which uses $A_{4;1}^{[0]}(1^-,2^+,3^-,4^+)$ as an input, and agrees
with the known result.)
In \sect{SixPointSection} we feed the five-point result
back into the recursion to construct the six-point result,
and following that, outline the construction of the seven-point
result.

Finally, we also need the three-point amplitudes, 
which vanish for real momenta, but are non-vanishing for 
generic complex momenta.  The one-loop three-vertex that we
need may be deduced from a one-loop splitting
amplitude~\cite{Neq4Oneloop}.

Let us start with the tree amplitudes.
The tree amplitudes that enter into our calculation are just 
the Parke-Taylor amplitudes~\cite{ParkeTaylor,BGSix,MPX},
\begin{eqnarray}
A_{n}^\tree(1^\pm,2^+, 3^+, \ldots, n^+) &=& 0 \,,\\
A_{n}^\tree(1^-,2^-, 3^+, \ldots, n^+) &=& 
 i {\spa1.2^4 \over \spa1.2 \spa2.3 \spa3.4 \cdots \spa{n}.1} \,.
\label{TwoMinusTree}
\end{eqnarray}
For some complex momenta, the three-point amplitudes
\begin{eqnarray}
A_{3}^\tree(1^-,2^-, 3^+) &=& 
 i {\spa1.2^4 \over \spa1.2 \spa2.3 \spa3.1} \,, \\
A_{3}^\tree(1^+,2^+, 3^-) &=& 
- i {\spb1.2^4 \over \spb1.2 \spb2.3 \spb3.1} \,,
\end{eqnarray}
are non-vanishing.  (They are vanishing for real momenta.)

The finite one-loop amplitudes that feed into our recursion are also
relatively simple.  The four-point finite amplitude with a single
negative-helicity leg is~\cite{BKStringBased,KunsztEtAl,TwoQuarkThreeGluon},
\begin{eqnarray}
A_{4;1}^{[0]}(1^-,2^+,3^+,4^+) = {i \cg \over 3} \,
     {\spa2.4 \spb2.4^3 \over \spb1.2 \spa2.3 \spa3.4 \spb4.1} \,.
\end{eqnarray}
Since the amplitude is entirely composed of rational functions,
the vertex is proportional to the amplitude
\begin{equation}
\Vertex_4(1^-, 2^+, 3^+, 4^+)=  {1\over i \cg} \, 
 A_{4;1}^{[0]} (1^-, 2^+, 3^+, 4^+) \,.
\end{equation}
We have removed an extra $i$ from this vertex (and all others in the
section) compared to the vertices $R_n$
appearing in \sect{DerivationSection}. As
mentioned at the end of that section, we wish to perform the recursion
directly on the finite parts $F_n^s$ defined in \eqn{An0}. 
For this reason, a factor of $i$ from the vertex is removed from the vertex, 
compared with the one that would be used for constructing the amplitudes $A_n$.

The five-point finite amplitudes are also rather simple.  
We will need the finite amplitude
\begin{eqnarray}
A_{5;1}^{[0]} (1^-, 2^+, 3^+, 4^+, 5^+)
& = & 
%%%%% begin : ampl5mpppp
i {\cg\over 3} \,  {1\over \spa3.4^2} 
\Biggl[-{\spb2.5^3 \over \spb1.2 \spb5.1}
       + {\spa1.4^3 \spb4.5 \spa3.5 \over \spa1.2 \spa2.3 \spa4.5^2}
       - {\spa1.3^3 \spb3.2 \spa4.2 \over \spa1.5 \spa5.4 \spa3.2^2} 
     \Biggr] \,. 
%%%%% end : ampl5mpppp
\label{mppppsimple} 
\nn
\end{eqnarray}
This amplitude, along with all the other one-loop five-gluon helicity
amplitudes, was first calculated using string-based
methods~\cite{GGGGG}.  Again because the amplitude is purely rational,
the vertex is proportional to the amplitude,
\begin{equation}
\Vertex_5(1^-, 2^+, 3^+, 4^+, 5^+) = {1\over i \cg} \, 
 A_{5;1}^{[0]} (1^-, 2^+, 3^+, 4^+, 5^+)  \,.
\end{equation}
It is worth noting that compact expressions now exist for the
$n$-point generalization of this
amplitude~\cite{OnShellRecurrenceI,Qpap}, 
$A_{n;1}^{[0]}(1^-,2^+,3^+,\ldots,n^+)$,
which agree numerically with
Mahlon's~\cite{Mahlon} original determination.

Using the decompositions (\ref{An0})--(\ref{AnQCD}), we express the divergent
four-point amplitudes $A_{4;1}^{[J]}(1^-,2^-, 3^+, 4^+)$ in
terms of the functions~\cite{BKStringBased,KunsztEtAl,TwoQuarkThreeGluon},
\begin{eqnarray}
V^g_4 &=& - {2\over \e^2}  \Biggl[ \biggl( {\mu^2 \over -s_{12}} \biggr)^\e
                + \biggl( {\mu^2 \over -s_{23}}\biggr)^\e \Biggr]
               + \ln^2\biggl({-s_{12}\over -s_{23}}\biggr) + \pi^2 - 
                 {\delta_R\ \over 3} \,, 
\label{Vg4}\\ 
V^f_4 &=& -{1\over \eps} \biggl({ \mu^2 \over -s_{23}} \biggr)^\e - 2 \, ,
\label{Vf4} \\
V^s_4 &=& - {V^f_4 \over 3} + {2\over 9} \,.
\label{Vs4}
\end{eqnarray}
In this case the finite parts are trivial,
\begin{equation}
F^f_4 = 0 \,,  \hskip 2 cm 
F^s_4 = 0 \,.
\label{mmppF}
\end{equation}
The version of dimensional regularization under consideration is
determined by the $\delta_R$ parameter; for the 't~Hooft-Veltman
scheme~\cite{HV} we take $\delta_R = 1$ while for the four-dimensional
helicity scheme~\cite{BKStringBased,FDH2} we take $\delta_R = 0$. 

\Eqn{mmppF} shows that all the rational terms for the $({-}{-}{+}{+})$
case are constant multiples of the tree amplitude, so they can easily
be absorbed into the $V^x$ terms.  
Because the tree amplitude obeys its own on-shell recursion
relation~\cite{BCFRecurrence}, it does not really matter whether we put
constant terms like the 2/9 term in $V_4^s$ into the $V$ or $F$
category, but we will be able to drop one recursive diagram by
assigning it to $V$.  In general, then, we define $V_n^s$ by,
\begin{equation}
V^s_n = - {V^f_n \over 3} + {2\over 9} \,.
\label{Vsn}
\end{equation}
Because $F_4^s$ vanishes, we take the loop vertex to also vanish,
\begin{equation}
\Vertex_4(1^-,2^-, 3^+, 4^+) = 0 \,.
\label{Ers4}
\end{equation}

For the divergent amplitudes with five or more legs, it is useful to
introduce a set of auxiliary functions~\cite{GGGGG},
\begin{eqnarray}
\Ll_0(r) &=& {\ln(r)\over 1-r}\,, \nn \\
\Ll_1(r) &=& {\ln(r)+1-r\over (1-r)^2}\,, \nn\\
\Ll_2(r) &=& {\ln(r)-(r-1/r)/2\over (1-r)^3} \,,
\label{Lsdef}
\end{eqnarray}
in which the pole at $r=1$ is removable. As discussed 
in \sect{DerivationSection}, we can therefore use the functions 
to construct the completed-cut terms, out of logarithms deduced 
from four-dimensional cuts. 

We shall be quoting the functional 
form of the cut-containing pieces in the Euclidean region; 
a discussion of analytic continuations to the physical region 
may be found in, for example, ref.~\cite{ZFourPartons}. 

We will also need the five-gluon amplitude
$A_{5;1}^{[J]}(1^-,2^-, 3^+, 4^+,5^+)$.  Ref.~\cite{GGGGG} gives us,
\begin{eqnarray}
\V^{g}_5 &=& -{1\over\e^2}\sum_{j=1}^5 \biggl({\mu^2\over -s_{j,j+1}}\biggr)^\e
          +\sum_{j=1}^5 \ln \biggl({-s_{j,j+1}  \over -s_{j+1,j+2}}\biggr)\,
                        \ln \biggl({-s_{j+2,j-2}\over -s_{j-2,j-1}}\biggr)
          +{5\over6}\pi^2 - {\delta_R\over3}\,, \label{Vg5} \hskip 1 cm  \\
\V^{f}_5&=&  -{1\over2 \e} \Biggl[  \biggl({\mu^2\over -s_{23}}\biggr)^{\e}
                         +\biggl({\mu^2\over -s_{51}}\biggr)^\e  \Biggr] - 2 
                  \,, \label{Vf5} \\
\V^{s}_5 &=& -{1\over3} \V^{f}_5 + {2\over9} \,,
  \label{Vs5}
\end{eqnarray}
for the functions
appearing in the decompositions (\ref{An0})--(\ref{AnQCD}).
(The $1/\e$ singularities differ from those in ref.~\cite{GGGGG} because
the amplitudes there were renormalized, whereas here we are using
unrenormalized amplitudes; the difference is
given simply by the renormalization subtraction~(\ref{MSbarsubtraction}).)
The finite parts of this amplitude are,
\begin{eqnarray}
\F^{f}_5 &=& -{1\over 2}
   {{\spa1.2}^2 \biggl(\spa2.3\spb3.4\spa4.1+\spa2.4\spb4.5\spa5.1\biggr) \over
    \spa2.3\spa3.4\spa4.5\spa5.1}
     {\Ll_0\biggl( {-s_{23}\over -s_{51}}\biggr)\over s_{51}} \,,
\label{Ff5} \\
\F^{s}_5 &=&
     - {1\over3}\F_5^{f} -{1\over 3}
   {\spb3.4\spa4.1\spa2.4\spb4.5
 \biggl(\spa2.3\spb3.4\spa4.1+\spa2.4\spb4.5\spa5.1 \biggr)\over\spa3.4\spa4.5}
     {\Ll_2\biggl( {-s_{23}\over -s_{51}} \biggr)\over s_{51}^3} \hskip 1.5 cm 
  \nn   \\
 &&\null 
  \hskip 2 cm  + \Remaining_5 \,,
\label{Fs5}
\end{eqnarray}
where 
\begin{eqnarray}
\Remaining_5 & =&
 -{1\over3}{\spa3.5{\spb3.5}^3\over\spb1.2\spb2.3\spa3.4\spa4.5\spb5.1}
     +{1\over3}{\spa1.2{\spb3.5}^2\over\spb2.3\spa3.4\spa4.5\spb5.1} \nn   \\
 && \null \hskip 2 cm 
     +{1\over6}{\spa1.2\spb3.4\spa4.1\spa2.4\spb4.5\over
                  s_{23}\spa3.4\spa4.5 s_{51}}\,.
\label{Remaining5}
\end{eqnarray} 
In \sect{FivePointSection}, we shall recompute the explicit rational
terms $\Remaining_5$.  The other pieces are all either trivial or
obtainable from four-dimensional unitarity.

Following the discussion of the previous section, we define a
recursion vertex (\ref{RationalDefinition}), 
composed of all the rational terms in
$F^s_5$,
\begin{equation}
\Vertex_5 (1^-,2^-, 3^+, 4^+, 5^+) = F^s_5 \Bigr|_\rat \,,
\end{equation}
including those contained in $\Ll_2$.  As in the
four-point case the effect of the $2/9$ term in $\V^s$
is trivial, so we do not need to make it part of
the recursion vertex.  

Now consider the known six-point results for $A_{6;1}^{[J]}(1^-,2^-,
3^+, 4^+, 5^+, 6^+)$.  Except for rational terms associated with the
$J=0$ scalar loop, these amplitudes were determined from the unitarity
method in refs.~\cite{Neq4Oneloop,Neq1Oneloop}, where the calculations
were performed for the more general $n$-point amplitudes with
two adjacent negative helicities.  
These results have recently been confirmed in refs.~\cite{BST,BBSTQCD}.  

From the results of ref.~\cite{Neq4Oneloop}, after setting $n=6$,
we have the $\NeqFour$ contribution,
\begin{equation}
V^g_6 =
%%%%% xbegin : Vg6
 \sum_{i=1}^{6} \Biggl[ -{ 1 \over \eps^2 }
 \Biggl({\mu^2  \over - s_{i,i+1}} \Biggr)^{\eps}
- \ln \biggl({ -s_{i,i+1}\over - s_{i,i+1,i+2} }\biggr)
  \ln \biggl({ - s_{i+1,i+2}\over - s_{i,i+1,i+2}} \biggr) \Biggr]
 + D_6 + L_6 + \pi^2 
%%%%% xend : Vg6
\,,
\label{UniversalFunc}
\end{equation}
where all indices are to be taken $\mod 6$ and
\begin{eqnarray}
D_{6} &=&
%%%%% xbegin : D6
 -\sum_{i=1}^{3} \li2  \Biggl( 1- {s_{i,i+1} s_{i+3,i+4}
        \over s_{i,i+1,i+2} s_{i-1,i,i+1} } \Biggr)
%%%%% xend : D6
\, , \nn \\
L_{6} &=&
%%%%% xbegin : L6
 -{1\over 4} \sum_{i=1}^6
  \ln \Biggl({ - s_{i,i+1,i+2} \over - s_{i+1,i+2,i+3} } \Biggr)
  \ln \Biggl({ -s_{i+1,i+2,i+3}\over - s_{i,i+1,i+2} } \Biggr)
%%%%% xend : L6
\, .
\end{eqnarray}
From ref.~\cite{Neq1Oneloop}, the $\NeqOne$ components are,
\begin{eqnarray}
\V^{f}_6 &=& 
%%%%% begin : Vf6
 -{1 \over 2\e} \Biggr(\Biggl({\mu^2 \over -s_{23} } \Biggr)^\e +
                           \Biggl({\mu^2 \over -s_{61} } \Biggr)^\e \Biggr)   
            - 2  
%%%%% end : Vf6
 \,, \\
\F^{f}_6 &=& 
%%%%% begin : Ff6
  {1\over 2 s_{12}} {\spa1.2^3 \over \spa2.3 \spa3.4 \spa4.5 \spa5.6 \spa6.1}
 \Biggl[
   {1\over s_{16}} \, 
     \Lz \biggl({s_{234} \over s_{16}} \biggr)  \Bigl(
       \tr_+[1 2 5 6] - \tr_+[1 2 (1+6) 5] \Bigr) \label{Ff6} \\
&& \null \hskip 5.5 cm
  + {1\over s_{234}} \Lz \biggl({s_{23} \over s_{234}} \biggr)
          \Bigl(\tr_+[1 2 3 4] - \tr_+[1 2 4 (2+3)] \Bigr)  \Biggr] 
%%%%% end : Ff6
\,, \hskip 1 cm \nn 
\end{eqnarray}
while the scalar loop contributions are,
\begin{eqnarray}
\V^{s}_6 &=& 
%%%%% begin : Vs6
 -{1\over 3} V^f_6 + {2\over 9}
%%%%% end : Vs6
\,,  \label{V6s}  \\
\F^{s}_6 &=& 
%%%%% begin : Fs6
 - {1\over 3 s_{12}^3} 
    {\spa1.2^3 \over \spa2.3 \spa3.4 \spa4.5 \spa5.6 \spa6.1} \nn \\
&& \null \hskip .4 cm  \times 
    \Biggl[
   {1\over s_{16}^3} \, 
     \Ll_2 \biggl({s_{234} \over s_{16} } \biggr)  \Bigl(
       (\tr_+[1 2 5 6])^2 \tr_+[1 2 (1+6) 5] 
  - \tr_+[1 2 5 6] (\tr_+[1 2 (1+6) 5])^2 \Bigr)  \nn \\
&& \hskip 1 cm \null 
  + {1\over s_{234}^3} \Ll_2 \biggl({s_{23} \over s_{234}} \biggr)
          \Bigl(\tr_+[1 2 4 (2+3)]  (\tr_+[1 2 3 4])^2 -
          (\tr_+[1 2 4 (2+3)])^2 \tr_+[1 2 3 4] \Bigr)  \Biggr]
         \cr
&& \null \hskip 3 cm 
 - {1\over 3} F^f_6  + \Remaining_6 
%%%%% end : Fs6
\,,
\label{Fs6}
\end{eqnarray}
where $\Remaining_6$ are the rational terms not contained in $\Ll_2$.
A key task of this paper will be to obtain an explicit formula for the
unknown rational function $\Remaining_6$.  We shall do so in
\sect{SixPointSection}.

Similarly, for the seven-point amplitude
$A_{7;1}^{[0]}(1^-,2^-,3^+,4^+,5^+,6^+,7^+)$, it is not difficult to
extract the functions from the $n$-point forms given in
refs.~\cite{Neq4Oneloop,Neq1Oneloop}.  Although we shall not discuss
the seven-point case in any detail, these functions enter into the computation
of the rational terms of this amplitude, as outlined in
\sect{SixPointSection}.

We still need a one-loop three-vertex for the recursion.  We determine
this vertex by inspecting the one-loop splitting amplitudes.  Because
the loop splitting amplitude with opposite on-shell helicities and a
scalar circulating in the loop vanishes (see \eqn{OneloopSplitExplicitmp}),
\begin{equation}
\Split_{-\lambda}^{[0]} (1^\pm, 2^\mp) = 0 \,,
\end{equation}
the corresponding three-vertex should also be taken to vanish,
\begin{equation}
\Vertex_3 (\hat 1^-,2^+, -\Kh_{12}^\pm) = 0\,.
\label{Ers3}
\end{equation}
For the cases where the two external lines have the same helicity, the
situation is much more subtle, with the appearance of `unreal
poles'~\cite{OnShellRecurrenceI,Qpap} and non-factorizing
contributions.  However, because we choose to shift the two 
negative-helicity legs, we do not encounter such vertices.

Finally, we need the rational functions $\CuthRat_n$
contained in the completed-cut
part defined in \eqns{CompletedCutDecomposition}{CuthRatDefinition}, 
for $n=5$, 6, and 7.  These functions
will be used to obtain the overlap contributions in
sections~\ref{FivePointSection} and \ref{SixPointSection}.  We easily
obtain these from the $\Ll_2$ terms by replacing the $\Ll_2$ function 
with its rational part, using \eqn{Lsdef}. 
For the five-point amplitude $A_{n;1}^{[0]}(1^-,2^-,
3^+, 4^+, 5^+)$, the explicit value is,
\begin{equation}
\CuthRat_5 =    
%%%%% begin : Cuth5Rat
 -{1\over 6} {s_{15} + s_{23} \over s_{23} s_{15} (s_{15} -  s_{23})^2}
   {\spb3.4\spa4.1\spa2.4\spb4.5
 \biggl(\spa2.3\spb3.4\spa4.1+\spa2.4\spb4.5\spa5.1 \biggr)\over\spa3.4\spa4.5}
%%%%% end : Cuth5Rat
    \,.
\label{R5Cut}
\end{equation}
For the six-point amplitude  $A_{6;1}^{[0]}(1^-,2^-, 3^+, 4^+, 5^+,6^+)$,
the rational function appearing in the completed-cut part is,
\begin{eqnarray}
\CuthRat_6 & = &
%%%%% begin : Cuth6Rat
  - {1\over 6} {1 \over  \spa2.3 
            \spa3.4 \spa4.5 \spa5.6 \spa6.1}  \nn \\
&& \hskip .5 cm \times 
\Biggl[ {1\over  s_{16} s_{234} } \,
          { s_{16}  + s_{234} \over 
                 (s_{16} - s_{234})^2} \,
    \spa2.5 \spb5.6 \spa6.1 \spa5.1 \sand2.{(3+4)}.5  \nn  \\
&& \null \hskip 2 cm \times
  \biggl( \spa2.5 \spb5.6 \spa6.1 + \sand2.{(3+4)}.5 \spa5.1  \biggr) \nn  \\
&& \null \hskip 1 cm  
  + {1\over s_{23} s_{234} } 
       {s_{234} + s_{23} \over (s_{234} - s_{23} )^2}
   \spa2.4 \sandpp4.{(5+6)}.1 \spa2.3 \spb3.4 \spa4.1   \nn \\
&& \null \hskip 2 cm \times
        \biggl(\spa2.3 \spb3.4 \spa4.1 + \spa2.4 \sandpp4.{(5+6)}.1
      \biggr)
          \Biggr] 
%%%%% end : Cuth6Rat
\,.
\label{R6Cut}
\end{eqnarray}
Similarly, for the seven-point amplitude $A_{7;1}^{[0]}(1^-,2^-, 3^+,
4^+, 5^+,6^+,7^+)$ which we briefly discuss in \sect{SixPointSection}, 
it is not difficult to extract the explicit
value of $\CuthRat_7$ from eq.~(7.1) of ref.~\cite{Neq1Oneloop},
although we shall not quote the result here.

%%%%%%%%%%%%%%%%%%%%%%%%%%%%%%%%%%%%%%%%%%%%%%%%%%%%

\section{Recomputation of Five-Gluon QCD Amplitudes}
\label{FivePointSection}

In this section we illustrate our method for determining loop
amplitudes, by recomputing the known five-gluon QCD amplitudes, given
the four-dimensional cut-constructible parts of the amplitudes.  
There are two independent helicity amplitudes,
$A_{5;1}^{[0]}(1^-,2^-,3^+, 4^+, 5^+)$ and $A_{5;1}^{[0]}(1^-,2^+,3^-,
4^+, 5^+)$.  We will discuss the first of these in some detail, and
merely summarize the calculation of the latter.  In both cases, we
correctly reproduce the results of ref.~\cite{GGGGG}.

Begin with $A_{5;1}^{[0]}(1^-,2^-,3^+, 4^+, 5^+)$. 
For this amplitude, use a $(1,2)$ shift,
\begin{eqnarray}
\lambda_1 &\rightarrow& \lambda_1\,, \nn\\
\tlambda_1 &\rightarrow& \tlambda_1 - z \tlambda_2 \,,\nn \\
\lambda_2 &\rightarrow& \lambda_2 + z \lambda_1 \,, \nn \\
\tlambda_2 &\rightarrow& \tlambda_2\,.
\label{SpinorShift12}
\end{eqnarray}
It is not difficult to verify that this shift has the required
property that the rational part of the cut term $\CuthRat_5$, given in
\eqn{R5Cut}, vanishes at large $z$, as required.

%%%%%%%%%%%%%%%%%%%%
%FIGURE
%
\begin{figure}[t]
\centerline{\epsfxsize 5.0 truein \epsfbox{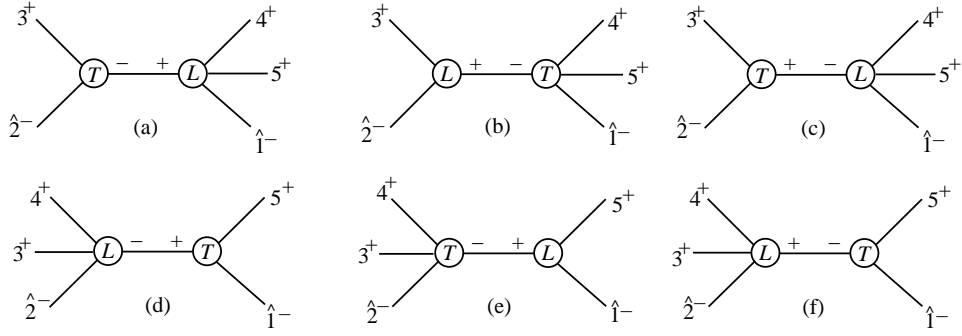}}
\caption{The recursive diagrams for computing the rational parts of
$A_{5;1}^{[0]}(1^-, 2^-, 3^+, 4^+, 5^+)$ with the shift of legs 1,2
given in \eqn{SpinorShift12}. `$T$' signifies a tree vertex and `$L$'
a loop vertex.}
\label{A5mmppp12Figure}
\end{figure}
%%%%%%%%%%%%%%%%%%%%%%

This shift yields a version of the
rational-recursion~(\ref{RationalRecursion}), where each term is
represented by one of the recursive diagrams depicted in
fig.~\ref{A5mmppp12Figure}. We have dropped diagrams with a
trivially vanishing tree amplitude.  Consider the first diagram in
\fig{A5mmppp12Figure},
\begin{equation}
D_5^{\rm (a)} = A_3^\tree(\hat 2^-, 3^+, -\Kh_{23}^-) 
           \times {i\over s_{23}}
           \times \Vertex_4 (\hat 1^-, \Kh_{23}^+, 4^+,
                 5^+) \,.
\end{equation}
It vanishes,
\begin{equation}
D_5^{\rm (a)} = 0 \,, 
\end{equation}
because~\cite{BCFRecurrence}
\begin{eqnarray}
 A_3^\tree(\hat 2^-, 3^+, -\Kh_{23}^-)  & \propto &
 \spash{\hat 2}.{\Kh_{23}}^3  \nn \\
&\propto & 
          \Bigl(\sand2.{(2+3)}.2 - {s_{23} \over \sand1.{(2+3)}.2}
              \sand1.{(2+3)}.2\Bigr)^3 \nn\\
& =& 0\,.
\end{eqnarray}
Diagram (b) also vanishes,
\begin{equation}
D_5^{\rm (b)} = 0 \,,
\end{equation}
because the loop three-vertex (\ref{Ers3}) vanishes.
Similarly, it is not difficult to show that diagrams (d) and (e) vanish,
\begin{equation}
D_5^{\rm (d)} = D_5^{\rm (e)} = 0 \,.
\end{equation}

We are left with just two direct-recursion diagrams.  Diagram (c) is given
by 
\begin{equation}
D_5^{\rm (c)} =  A_3^\tree(\hat 2^-, 3^+, -\Kh_{23}^+) 
          \times  {i\over s_{23}} \times 
                 \Vertex_4 (\hat 1^-, \Kh_{23}^-, 4^+, 5^+) \,.
\end{equation}
As we saw in \eqn{Ers4}, the loop vertex vanishes, and
so 
\begin{equation}
D_5^{\rm (c)} = 0 \,.
\end{equation}
The last diagram is,
\begin{eqnarray}
D_5^{\rm (f)} 
& =& A_3^\tree(5^+,\hat 1^-, -\Kh_{51}^-) \times {i\over s_{51}} \times
                 \Vertex_4 (\hat 2^-, 3^+, 4^+, \Kh_{51}^+) \nn \\
 &=& - {1\over 3} {\spash{\hat 1}.{(-\Kh_{51})}^3 \over \spash{5}.{\hat 1} 
           \spash{(-\Kh_{51})}.5 } \, {1\over s_{51}} \,
       {\spash3.{\Kh_{51}} \spbsh3.{\Kh_{51}}^3 \over 
       \spbsh{\hat 2}.3 \spa3.4 \spash4.{\Kh_{15}} \spbsh{\Kh_{51}}.{\hat 2} }
                     \nn\\
 &=&  {1\over 3} {\sand{1}.{5}.2 ^3 \over \spa{5}.{1} 
           \sand5.{1}.2}  \, {1\over \spa5.1 \spb1.5 \sand1.5.2^2} \,
       {\sand3.{4}.2 \sand1.{5}.3^3 \over 
       \spb2.3 \spa3.4 \sand4.{3}.2 \sand1.5.2} \nn \\
 & = & 
%%%%% begin : Df5
- {1 \over 3} {\spb2.4\spb3.5^3\over \spa3.4\spb1.2\spb1.5\spb2.3^2} \,.
%%%%% end : Df5
\end{eqnarray}
This direct-recursion diagram is the only one that does not vanish.

%%%%%%%%%%%%%%%%%%%%
%FIGURE
%
\begin{figure}[t]
\centerline{\epsfxsize 4 truein \epsfbox{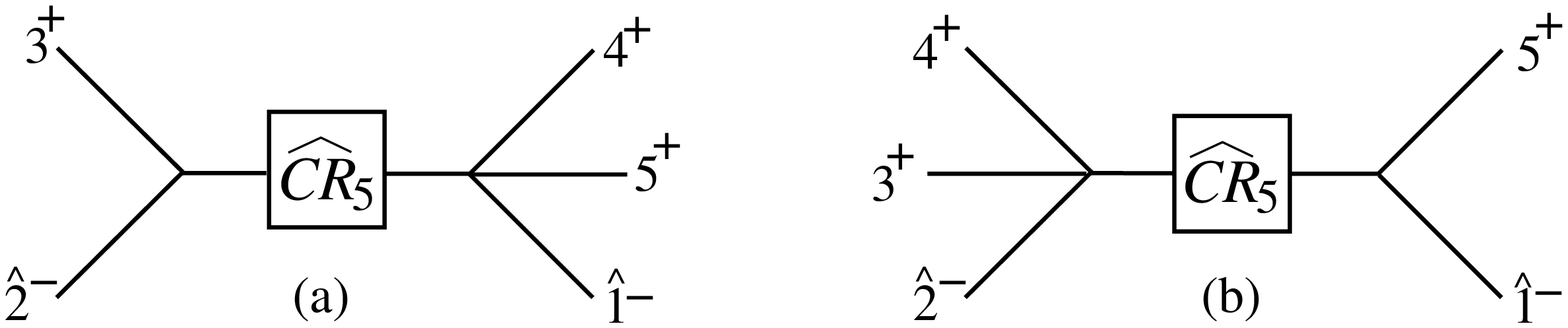}}
\caption{The five-point overlap diagrams using the (1,2) shift given in 
  \eqn{SpinorShift12}.}
\label{Overlap5Figure}
\end{figure}
%%%%%%%%%%%%%%%%%%%%%%

Next we must evaluate the overlap contributions 
from \eqn{BasicBootstrapEquation},
depicted in \fig{Overlap5Figure}.  We start from the
rational parts of the cut contributions, $\CuthRat_5$ as given
in \eqn{R5Cut}.  
Applying the shift~(\ref{SpinorShift12}) to these contributions, 
we have,
\begin{eqnarray}
\CuthRat_5(z) & = &
%%%%% begin : Cuth5RatZ
- {1\over 6} {\spb3.4\spa4.1(\spa2.4 + z \spa1.4) \spb4.5
    \over\spa3.4\spa4.5} \nn \\
&& \null \hskip .3cm \times
 \biggl((\spa2.3+ z \spa1.3)  \spb3.4 \spa4.1+ (\spa2.4 + z \spa1.4)
            \spb4.5\spa5.1 \biggr)   \\
&& \null \hskip .3cm \times
    {s_{51} + s_{23}  -z \sand1.5.2  + z \sand1.3.2 
         \over (\spa2.3 + z \spa1.3) \spb3.2 \spa1.5 (\spb5.1 -z \spb5.2)
           (s_{15} - s_{23} - z \sand1.{(5+3)}.2)^2 }
%%%%% end : Cuth5RatZ
 \,. \nn
\end{eqnarray}
The residues of $\CuthRat_5(z)/z$ that we need to evaluate are
located at the values of $z$,
\begin{equation}
z^{\rm (a)} =  - {\spa2.3 \over \spa1.3}\,,  \hskip 2 cm 
z^{\rm (b)} =    {\spb1.5 \over \spb2.5} \,, \hskip 2 cm 
\end{equation}
corresponding to the two overlap diagrams in \fig{Overlap5Figure}. 
Evaluating the residue corresponding to the first of these overlap diagrams 
is very simple and gives, 
\begin{eqnarray}
O_5^{\rm (a)} &=&
%%%%% begin : O5a
- {1\over6} {\spa1.2^2\spa1.4\spb3.4
           \over \spa1.5\spa2.3\spa3.4\spa4.5\spb2.3} 
%%%%% end : O5a
 \,.
\end{eqnarray}
Similarly, the overlap diagram (b) gives 
\begin{eqnarray}
O_5^{\rm (b)} &=& 
%%%%% begin : O6b
{1\over6} {\spa1.4\spb3.4\spb3.5 \bigl(\spa1.4\spb3.4-\spa1.5\spb3.5\bigr)
            \over\spa1.5\spa3.4\spa4.5\spb1.5\spb2.3^2} 
%%%%% end : O6b
\,.
\end{eqnarray}

Summing over the non-vanishing diagrammatic contributions, we get
the simple result, 
\begin{eqnarray}
\Remaining_5 &=& D_5^{\rm (f)} + O_5^{\rm (a)} + O_5^{\rm (b)} \nn \\
& = &
%%%%% begin : DiagramSum5
 {1\over6} \Biggl(
 - {2 \spb2.4 \spb3.5^3\over \spa3.4 \spb1.2 \spb1.5 \spb2.3^2}
-{\spa1.2^2 \spa1.4 \spb3.4\over \spa1.5 \spa2.3 \spa3.4 \spa4.5 
    \spb2.3}
  \nn \\
&& \null \hskip .8 cm 
 + {\spa1.4^2 \spb3.4^2 \spb3.5\over\spa1.5 \spa3.4 \spa4.5 \spb1.5 \spb2.3^2}
 -{\spa1.4 \spb3.4 \spb3.5^2\over\spa3.4 \spa4.5 \spb1.5 
   \spb2.3^2}
\Biggr)  
%%%%% end : DiagramSum5
\,. \hskip 1.5 cm 
\end{eqnarray}
With a few spinor manipulations this result can be brought into
manifest agreement with the known
result (\ref{Remaining5}).  Thus, we have correctly reproduced the rational
parts of $A_{5;1}^{[0]}(1^-, 2^-, 3^+,4^+, 5^+)$, without
performing any loop integrals.

We have also verified that our method properly reproduces
$A_{5;1}^{[0]}(1^-, 2^+, 3^-, 4^+, 5^+)$, computed in
ref.~\cite{GGGGG}.  As for the previous case, we choose the
negative-helicity legs as the shifted ones, {\it i.e.} we use 
a $(1,3)$ shift,
\begin{eqnarray}
\lambda_1 &\rightarrow& \lambda_1\,, \nn\\
\tlambda_1 &\rightarrow& \tlambda_1 - z \tlambda_3 \,,\nn \\
\lambda_3 &\rightarrow& \lambda_3 + z \lambda_1 \,, \nn \\
\tlambda_3 &\rightarrow& \tlambda_3\,.
\label{SpinorShift13}
\end{eqnarray}
For this computation, we completed the cut terms
with the $\Ls_1$ function 
of ref.~\cite{GGGGG}, along with the $\Ll_1$ and $\Ll_2$ functions 
also used there,
\begin{eqnarray}
\Cuth_5 &&=
      - {{\spa1.2} {\spa2.3} {\spa3.4}
          {\spa4.1}^2 {\spb2.4}^2
         \over {\spa4.5} {\spa5.1} {\spa2.4}^2}\,
     {2 \, \Ls_1\Bigl( {-s_{23}\over -s_{51}},\,{-s_{34}\over -s_{51}}\Bigr)
      + \Ll_1\Bigl( {-s_{23}\over -s_{51}}\Bigr)
      + \Ll_1\Bigl( {-s_{34}\over -s_{51}}\Bigr)  \over s_{51}^2} \nonumber\\
&& \hskip 4mm
    \null  + {{\spa3.2} {\spa2.1} {\spa1.5}
          {\spa5.3}^2 {\spb2.5}^2
         \over {\spa5.4} {\spa4.3} {\spa2.5}^2}\,
        {2 \, \Ls_1\Bigl( {-s_{12}\over -s_{34}},\,{-s_{51}\over -s_{34}}\Bigr)
        + \Ll_1\Bigl( {-s_{12}\over -s_{34}}\Bigr)
        + \Ll_1\Bigl( {-s_{51}\over -s_{34}}\Bigr)  \over s_{34}^2} \nonumber\\
&& \hskip 4mm
    \null  +{2\over 3} {{\spa2.3}^2 {\spa4.1}^3 {\spb2.4}^3
          \over {\spa4.5} {\spa5.1} {\spa2.4}}
          {\Ll_2\Bigl( {-s_{23}\over -s_{51}}\Bigr)  \over s_{51}^3}
      -{2\over 3} {{\spa2.1}^2 {\spa5.3}^3 {\spb2.5}^3
          \over {\spa5.4} {\spa4.3} {\spa2.5}}
          {\Ll_2\Bigl( {-s_{12}\over -s_{34}}\Bigr)  \over s_{34}^3} 
\label{NonAdjacentCuth5}\\
&& \hskip 4mm
   \null + {\Ll_2\Bigl( {-s_{34}\over -s_{51}}\Bigr)\over s_{51}^3}\,
     \biggl( {1\over3} { \spa1.3\spb2.4\spb2.5
     (\spa1.5 \spb5.2 \spa2.3-\spa3.4 \spb4.2 \spa2.1) \over \spa4.5}
  \nonumber\\
&&\hskip 4mm
    \hphantom{ +\Ll_2\Bigl( {-s_{34}\over -s_{51}}\Bigr) () }
 \null   +{2\over 3} {{\spa1.2}^2{\spa3.4}^2\spa4.1{\spb2.4}^3
            \over \spa4.5\spa5.1\spa2.4}
    -{2\over 3} {{\spa3.2}^2{\spa1.5}^2\spa5.3{\spb2.5}^3
            \over \spa5.4\spa4.3\spa2.5}  \biggr) \nonumber\\
&&\hskip 4mm
  \null +{1\over 6} {{\spa1.3}^3
        \bigl( \spa1.5\spb5.2\spa2.3 - \spa3.4\spb4.2\spa2.1\bigr)
          \over \spa1.2\spa2.3\spa3.4\spa4.5\spa5.1}\,
            {\Ll_0\Bigl( {-s_{34}\over -s_{51}}\Bigr)\over s_{51}}
            \ .\nonumber
\end{eqnarray}
This function satisfies $\Cuth_5(z) \rightarrow 0$ as
$z\rightarrow\infty$, thanks to cancellations between the
polylogarithmic or logarithmic functions and the rational terms.  The
attentive reader will note that this function has spurious double
poles involving non-adjacent legs in the color ordering, {\it e.g.}
$1/\spa2.4^2$.  These poles do not invalidate the calculation with a
$(1,3)$ shift, because they acquire no $z$ dependence, and hence
produce no poles in $z$ at spurious locations.  The spurious
singularities cancel in the complete answer for $\F^s$, not only at
order $1/\spa2.4^2$, but also at order $1/\spa2.4$, even for complex
momenta.  (The completed-cut term $\Cuth_5$ given
here does contain a $1/\spa2.4$ pole for complex momenta, but it is
cancelled in the full answer by the additional diagrammatic terms in
our construction.)

%%%%%%%%%%%%%%%%%%%%%%%%%%%%%%%%%%%%%%%%%%%%%%%%%%%%

\section{Six- and Seven-Point QCD Amplitudes}
\label{SixPointSection}

%%%%%%%%%%%%%%%%%%%%
%FIGURE
%
\begin{figure}[t]
\centerline{\epsfxsize 5.0 truein \epsfbox{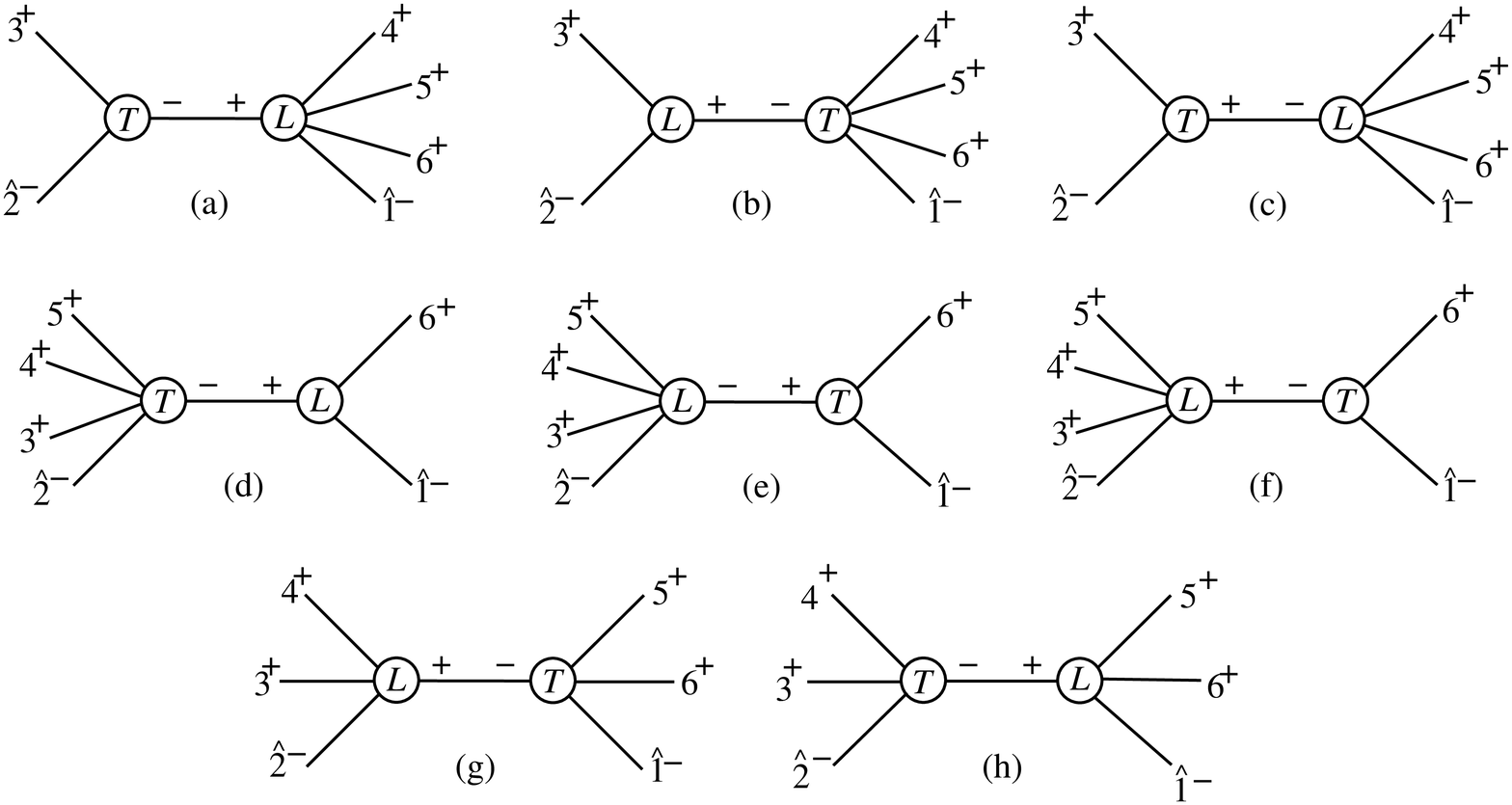}}
\caption{The recursive diagrams for $A_{6;1}^{[0]}(1^-, 2^-, 3^+, 4^+,
5^+,6^+)$, using the $(1,2)$ shift. Diagrams (a), (b), (d) and (e)
vanish.}
\label{A6mmppp12Figure}
\end{figure}
%%%%%%%%%%%%%%%%%%%%%%

In this section we describe the computations of the unknown rational
functions for the six- and seven-point helicity amplitudes
$A_{6;1}^{[0]}(1^-,2^-,3^+,4^+,5^+,6^+)$ and
$A_{7;1}^{[0]}(1^-,2^-,3^+,4^+,5^+,6^+,7^+)$.  Were one to attempt the
calculation by traditional means, one would encounter large numbers of
Feynman diagrams.  The total number of one-loop diagrams for the
six-gluon process in QCD is 10,860 (including gluon, ghost and fermion
loops, but dropping those which vanish trivially in dimensional
regularization).  For seven and eight external gluons the numbers grow
to 168,925 and 3,017,490 respectively.  But the number of diagrams
only hints at the full complexity of the calculation, because it does
not take into account the explosion of terms resulting from tensor
integral reductions, which is what renders a brute-force Feynman
diagram computation impractical even on a modern computer.  The
approach we take in the present paper avoids this explosion by
focusing on analytic properties that all amplitudes must satisfy.  We
computed the logarithmic terms in the amplitudes long
ago~\cite{Neq4Oneloop,Neq1Oneloop}, using the unitarity-based method.
Here we complete the QCD calculation by computing the rational terms
in the scalar contributions, namely \eqn{NearestNeighborn} for
$n=6,7$.  For the six-point case we present a compact analytical
expression.

First consider the six-point amplitude
$A_{6;1}^{[0]}(1^-,2^-,3^+,4^+,5^+,6^+)$.  To obtain the
rational terms of the six-point amplitude, $\Remaining_6$ 
in \eqn{Fs6}, we first
evaluate the recursive diagrams shown in \fig{A6mmppp12Figure},
corresponding to the terms in the recursion (\ref{RationalRecursion}).
We again shift the two negative-helicity legs, using
\eqn{SpinorShift12}.  Following a similar discussion as for the
five-point amplitude, it is not difficult to show that diagrams (a),
(b), (d), and (e) vanish,
\begin{eqnarray}
D_6^{\rm (a)} = D_6^{\rm (b)} = D_6^{\rm (d)} = D_6^{\rm (e)} = 0 \,.
\end{eqnarray}
Four diagrams remain to be evaluated,  
\begin{eqnarray}
D_6 & = &  D_6^{\rm (c)} + D_6^{\rm (f)} + D_6^{\rm (g)} + D_6^{\rm (h)} \nn \\
  & = & A_3^\tree(\hat 2^-, 3^+, -\Kh_{23}^+) 
                \times {i\over s_{23}} \times
            \Vertex_5(\hat 1^-, \Kh_{23}^-, 4^+, 5^+, 6^+)   \nn \\
 && \null
          + \Vertex_5(\hat 2^-, 3^+, 4^+, 5^+, \Kh_{61}^+)
                \times {i\over s_{61}} \times
                  A_3^\tree(6^+, \hat 1^-, -\Kh_{61}^-)  \nn \\
 && \null
         + \Vertex_4(\hat 2^-, 3^+, 4^+, -\Kh_{234}^+)
                \times {i\over s_{234}} \times
           A_4^\tree(\hat 1^-, \Kh_{234}^-, 5^+, 6^+)  \nn \\
 && \null
         + A_4^\tree( -\Kh_{234}^-, \hat 2^-, 3^+, 4^+)
                \times {i\over s_{234}} \times
           \Vertex_4(\hat 1^-, \Kh_{234}^+, 5^+, 6^+)  \,.
\end{eqnarray}
We do not present a detailed evaluation of these 
diagrams, because it is similar to the five-point evaluation
discussed in the previous section.

%%%%%%%%%%%%%%%%%%%%
%FIGURE
%
\begin{figure}[t]
\centerline{\epsfxsize 5.4 truein \epsfbox{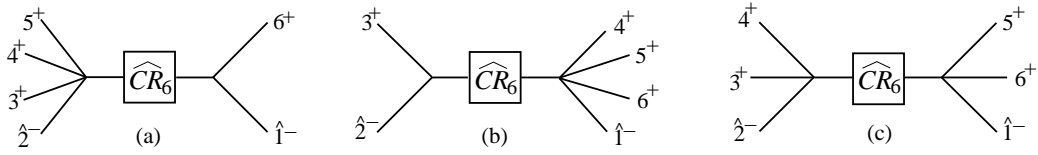}}
\caption{The overlap diagrams for $A_{6;1}^{[0]}(1^-,2^-, 3^+, 4^+, 5^+,6^+)$, 
using the (1,2) shift.}
\label{Overlap6Figure}
\end{figure}
%%%%%%%%%%%%%%%%%%%%%%

Next consider the overlap contributions of \eqn{BasicBootstrapEquation} 
displayed in 
\fig{Overlap6Figure}.  These are determined by
evaluating the residues of the rational part of the cut contributions
given in \eqn{R6Cut} after performing the shift (\ref{SpinorShift12}).
The residues of ${\CuthRat_6(z)/ z}$ that need to be computed are
at the values of $z$ located at,
\begin{equation}
z^{\rm (a)} = {\spb1.6 \over \spb2.6} \,, \hskip 2 cm 
z^{\rm (b)} = -{\spa2.3 \over \spa1.3} \,, \hskip 2 cm 
z^{\rm (c)} = -{s_{234} \over \sand1.{(3+4)}.2} \,,
\end{equation}
corresponding to the three overlap diagrams in \fig{Overlap6Figure}.
It is straightforward to evaluate the residues since we encounter
only simple poles.
For diagram (a) of \fig{Overlap6Figure}, for example, we have,
\begin{equation}
O_6^{\rm (a)} =
- {1 \over 6} { \spa1.5 \spb5.6 s_{612} \spab5.{(1+2)}.6
   ( 2 \spa1.5 s_{612} - \sandmp1.{2 (1+6)}.5 ) 
  \over \spa3.4 \spa4.5 \spa 5.6 \spa 6.1 \spb6.1
      \spab3.{(1+2)}.6 {\spab5.{(3+4)}.2}^2 } \,,
\label{Oasimp}
\end{equation}
which has been simplified using spinor-product identities.
Similarly, it is not difficult to evaluate the other two residues,
corresponding to diagrams (b) and (c) in \fig{Overlap6Figure}.

The sum of the non-vanishing recursive and overlap diagrams 
for $\Remaining_6$ in \eqn{Fs6} is given by
\begin{equation}
\Remaining_6 = D_6^{\rm (c)} + D_6^{\rm (f)} + D_6^{\rm (g)} + D_6^{\rm (h)} 
           + O_6^{\rm (a)} + O_6^{\rm (b)} + O_6^{\rm (c)} \,.
\end{equation}
One of the terms in the expression for the overlap diagram
in the $\spa2.3$ channel, $O^{\rm (b)}$, 
contains an unusual factor in its denominator, namely the 
square of the quantity
\begin{equation}
 \sandmp1.{5 (1+2)}.3 + \spa1.3 s_{56}
= \sandmp1.{(2+3)4}.3 - \sandmp1.{6(1+2)}.3  \,.
 \label{unusualdenom} 
\end{equation}
However, the recursive diagram in the $\spa2.3$ channel,
$D^{\rm (c)}$, shown in \fig{A6mmppp12Figure}(c), also contains a term
with this behavior, and the two terms cancel against each other.
After simplifying the sum over diagrams we obtain, 
\begin{eqnarray}
\Remaining_6 
& = &
%%%%% begin : rational6
{1 \over 6} \Biggl\{ 
- 2 { \spa3.5 \spb3.5 
    \spab4.{(1+2)}.3 \spab4.{(1+2)}.6 \spab5.{(1+2)}.6
  \over \spb1.2 {\spa3.4}^2 {\spa4.5}^2 \spb6.1
         \spab5.{(3+4)}.2 \spab6.{(1+2)}.3 } 
  \nn \\
&& \null \hskip .5 cm 
- 2 { \spa3.5 \spb3.6 {\spab4.{(1+2)}.6}^2 
 \over \spb1.2 {\spa3.4}^2 {\spa4.5}^2 \spb6.1
        \spab5.{(3+4)}.2 } 
  \nn \\
&& \null \hskip .5 cm 
+ 2 {\spa1.2 \spa2.4 \spa3.5 
       {\spb3.5}^2 \spb5.6 \spab5.{(1+2)}.6
  \over {\spa3.4}^2 \spa4.5 \spb6.1
      \spab2.{(1+6)}.5 \spab5.{(3+4)}.2 \spab6.{(1+2)}.3 }
  \nn \\
&& \null \hskip .5 cm 
+ 2 { {\spa1.2}^2 {\spb3.5}^2
    ( \sandmp5.{(3+4) \, 2}.1 + \sandmp5.{3 \, 5}.1 )
   \over \spa3.4 \spa4.5 \spa6.1 
       \spab2.{(1+6)}.5 \spab5.{(3+4)}.2 \spab6.{(1+2)}.3 }
  \nn \\
&& \null \hskip .5 cm 
- { {\spa1.2}^3 \spa3.5 \spb4.6 \spb5.6
   \over \spa2.3 \spa3.4 \spa4.5 \spa5.6 
        \spab1.{(2+3)}.4 \spab3.{(1+2)}.6 }
+ 2 { {\spb3.6}^3 
   \over \spb1.2 \spb2.3 {\spa4.5}^2 \spb6.1 }
  \nn \\
&& \null \hskip .5 cm 
- { \spb5.6 {\spab5.{(1+2)}.6}^2
    ( 2 \sandmp4.{(3+5)(1+2)}.5 + \spa1.2 \spb1.2 \spa4.5 )
   \over \spb1.2 \spa3.4 {\spa4.5}^2 \spa5.6 \spb6.1
        \spab3.{(1+2)}.6 \spab5.{(3+4)}.2 } 
  \nn \\
&& \null \hskip .5 cm 
+ 2 { {\spa1.5}^2 {\spb3.4}^2 \spb5.6
   ( \spa1.6 \spb3.4 \spa4.5 - \spab1.{(2+4)}.3 \spa5.6 )
   \over \spb2.3 \spa4.5 {\spa5.6}^2 s_{234}
        \spab1.{(2+3)}.4 \spab5.{(3+4)}.2 }
  \nn \\
&& \null \hskip .5 cm 
- { \spa1.2 \spa1.5 \spb3.4 \spb5.6 
    \sandmp1.{(5+6)(3+4)}.5
   \over \spa3.4 \spa4.5 \spa5.6 s_{234}
    \spab1.{(2+3)}.4 \spab5.{(3+4)}.2 }
  \nn \\
&& \null \hskip .5 cm 
+ 2 { \spa3.5 {\spab1.{(2+4)}.3}^3
   \over \spb2.3 \spa3.4 \spa4.5 \spa5.6 \spa6.1 s_{234}
       \spab5.{(3+4)}.2 }
  \nn \\
&& \null \hskip .5 cm 
- { \spa1.2 \spab1.{(2+4)}.3 
       ( 2 \spab1.{(2+4)}.3 + \spab1.{4}.3 )
   \over \spb2.3 \spa3.4 \spa4.5 \spa5.6 \spa6.1 s_{234} }
  \nn \\
&& \null \hskip .5 cm 
+ 2 { {\spa1.2}^3 {\spb4.6}^2 \spab5.{(4+6)}.5
   \over \spa2.3 \spa4.5 \spa5.6 s_{123}
       \spab1.{(2+3)}.4 \spab3.{(1+2)}.6 }
  \nn \\
&& \null \hskip .5 cm 
+ 2 { {\spa1.2}^3 {\spb3.5}^2 \spab4.{(3+5)}.4
   \over \spa3.4 \spa4.5 \spa6.1 s_{612}
       \spab2.{(1+6)}.5 \spab6.{(1+2)}.3 }
  \nn \\
&& \null \hskip .5 cm 
- { {\spa1.2}^2
   \over \spa2.3 \spa3.4 \spa4.5 \spa5.6 \spa6.1 }
    \Biggl[ { \spab1.{4}.3 \over \spb2.3 }
          + { \spab2.{5}.6 \over \spb6.1 } \Biggr]
\Biggl\}
%%%%% end : rational6
\,,
\label{rational6}
\end{eqnarray}
determining the previously unknown rational function
in \eqn{Fs6}.

As in the past, for both theoretical and practical reasons, it 
probably will be important to have the simplest representations of
amplitudes.  It is likely that the result in \eqn{rational6} can be
simplified even further.  For example, individual terms contain
spurious singularities due to the following denominator factors:
$\spab1.{(2+3)}.4$, $\spab2.{(1+6)}.5$, $\spab3.{(1+2)}.6$,
$\spab5.{(3+4)}.2$, and $\spab6.{(1+2)}.3$.  We have checked
numerically that these singularities cancel between different terms,
so that the full expression is non-singular.  Nevertheless, one might
expect a simpler form with fewer such cancellations.  Some such
cancellations have already been carried out to arrive at
\eqn{rational6}; besides the cancellation involving the unusual
factor~(\ref{unusualdenom}), individual recursive and overlap diagrams
also contained factors of $1/{\spab5.{(3+4)}.2}^2$, as in \eqn{Oasimp}.
These squared factors
were eliminated using spinor-product identities to combine and
rearrange terms.  

Removing {\it all} of the factors of the form
$\spab{a}.{(b+c)}.d$ is unlikely to be desirable; after all, the
simple forms of tree amplitudes found in
refs.~\cite{NeqFourSevenPoint,NeqFourNMHV,RSVNewTree,%
BCFRecurrence,TreeRecurResults} are simpler than previous forms
precisely {\it because} of the presence of some such denominators.
However, the denominators present in \eqn{rational6} occur in a rather
asymmetric fashion, preventing the flip symmetry of the expression ---
symmetry under
\begin{equation}
 1 \lr 2, \quad 3 \lr 6, \quad 4 \lr 5
\label{flipmmpppp}
\end{equation}
--- from being manifest.
Indeed, the factor $\spab5.{(3+4)}.2$ occurs in the denominator
many times, yet its image under the flip~(\ref{flipmmpppp}),
$\spab4.{(5+6)}.1$ never appears. 

We have carried out several numerical checks of \eqn{Fs6}, after
inserting into it the value of $\Remaining_6$ from \eqn{rational6}.
Besides verifying the absence of any spurious singularities, we
checked that the amplitude satisfies the non-manifest flip
symmetry~(\ref{flipmmpppp}).  We checked the multi-particle
factorization in the $s_{123}$ and $s_{234}$ channels. (The $s_{612}$
channel is related trivially to the $s_{123}$ channel
by the flip symmetry.)
We also confirmed the proper collinear behavior, for real momenta, in
all the independent channels $s_{12}$, $s_{23}$, $s_{34}$ and
$s_{45}$.  These checks leave little doubt that \eqn{rational6} is the
correct expression.

%%%%%%%%%%%%%%%%%%%%
%FIGURE
%
\begin{figure}[t]
\centerline{\epsfxsize 6. truein \epsfbox{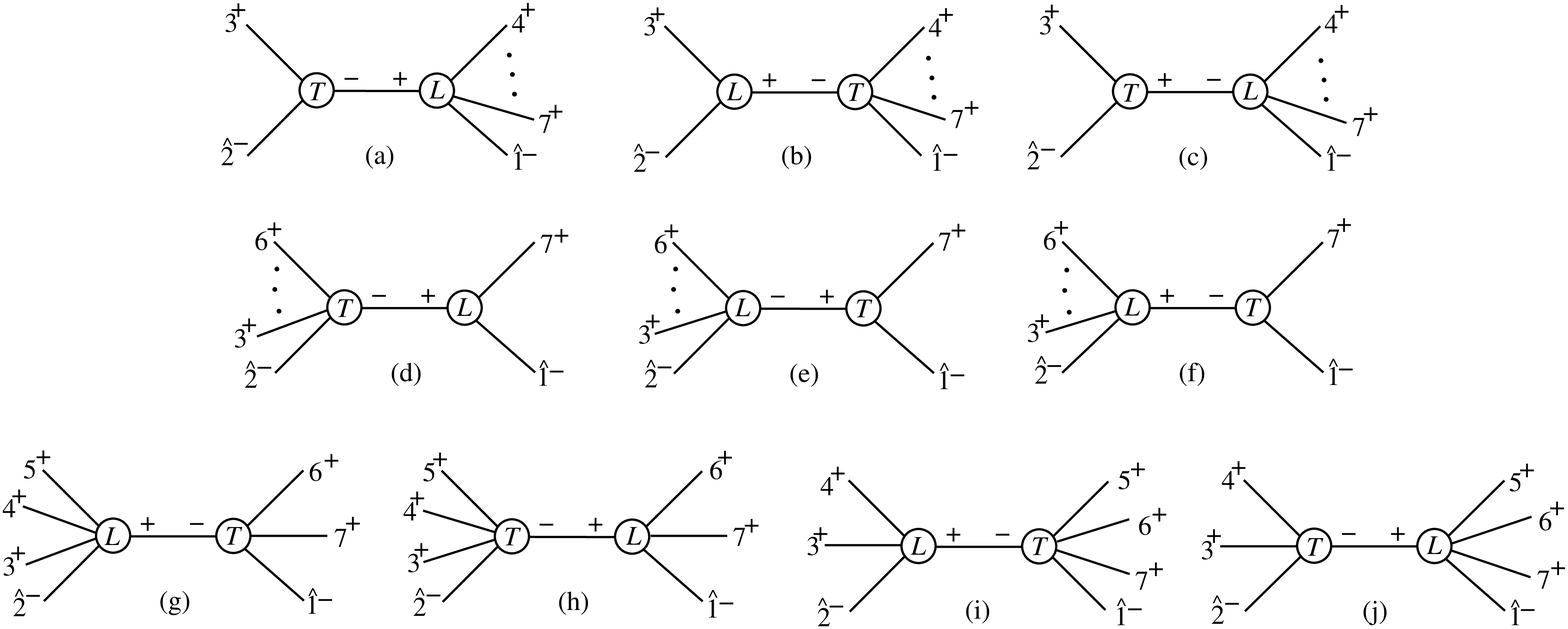}}
\caption{The recursive diagrams for the 
amplitude $A_{7;1}^{[0]}(1^-, 2^-, 3^+, 4^+, 5^+,6^+,7^+)$, using the
(1,2) shift. Diagrams (a), (b), (d) and (e) vanish. }
\label{A7mmpppp12Figure}
\end{figure}
%%%%%%%%%%%%%%%%%%%%%%

%%%%%%%%%%%%%%%%%%%%
%FIGURE
%
\begin{figure}[t]
\centerline{\epsfxsize 6. truein \epsfbox{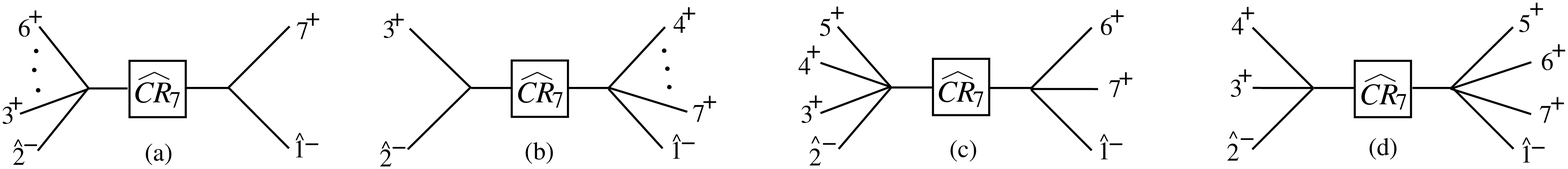}}
\caption{The overlap diagrams for the seven-point amplitude.}
\label{Overlap7Figure}
\end{figure}
%%%%%%%%%%%%%%%%%%%%%%

Using our expression for the six-point amplitude as well as the
seven-point cut terms from ref.~\cite{Neq1Oneloop} as input, it is then
straightforward to evaluate the complete seven-point scalar loop
amplitude, $A_{7;1}^{[0]}(1^-,2^-,3^+,4^+,5^+,6^+,7^+)$.  The needed
recursive diagrams are displayed in \fig{A7mmpppp12Figure}, while the
overlap diagrams are given in \fig{Overlap7Figure}.  We have evaluated
these diagrams and have confirmed that the resulting expression for
the amplitude has all the proper factorization properties in real
momenta, and that all spurious singularities cancel. Combining this
result with the known $\NeqFour$~\cite{Neq4Oneloop} and
$\NeqOne$~\cite{Neq1Oneloop} supersymmetric amplitudes, yields
a complete solution for the seven-gluon QCD amplitude with the same
helicity configuration.
Although its construction is entirely straightforward,
and parallels the six-point case, the seven-point result is rather 
lengthy, so we refrain from presenting it here.
Since the original version of this paper appeared, the result has 
been presented~\cite{FordeKosower}, as a member of the infinite series 
of $n$-point amplitudes $A_{n;1}^{\rm QCD}(1^-,2^-,3^+,4^+,\ldots,n^+)$,
constructed using the methods of the present paper.

These examples demonstrate the power of the factorization-bootstrap
approach, systematized here, as a complement to the unitarity-based
method, for evaluating complete QCD amplitudes, including purely
rational parts.  The required diagrams are surprisingly simple to
evaluate, not really more involved than tree-level diagrams.  It is
striking that what had previously been the most difficult part of a
one-loop QCD calculation has been reduced to a simple computation.

%%%%%%%%%%%%%%%%%%%%%%%%%%%%%%%%%%%%%%%%%%%%%%%%%%%%%

\section{Conclusions}

In this paper we presented a new method for computing the rational
functions in non-supersymmetric gauge theory loop amplitudes.  The
unitarity
method~\cite{Neq4Oneloop,Neq1Oneloop,Massive,UnitarityMachinery,BCFII}
has already proven itself to be an effective means for obtaining the
cut-containing terms in amplitutes, so we may rely on this approach
for obtaining such terms.  To obtain the rational terms we took a
recursive approach, systematizing an earlier unitarity-factorization
bootstrap~\cite{ZFourPartons}.

Our systematic loop-level recursion uses the proof of tree-level
on-shell recursion relations by Britto, Cachazo, Feng and
Witten~\cite{BCFW} as a starting point.  There are, however, a number
of issues and subtleties that arise, which are not present at tree level.
The most obvious issue is that the tree-level proof
relies on the amplitudes having only simple poles and no branch cuts;
loop amplitudes in general contain branch cuts.  Furthermore, as we
have already discussed in refs.~\cite{OnShellRecurrenceI,Qpap}, there
are subtleties resulting from the differences of one-loop
factorizations in complex momenta as compared to those in real
momenta.  These differences have important effects, unlike the
tree-level case.  At loop-level there are also spurious poles present,
which would interfere with a naive recursion on the rational terms.
In this paper we showed how to overcome these potential difficulties.

As an illustrative example of our approach, we described in some
detail a computation of the rational terms appearing in the five-gluon
QCD amplitudes with nearest-neighboring negative helicities in the color
ordering, reproducing the results~\cite{GGGGG} of the string-based
calculation of the same amplitudes.  Although we did not
describe it in any detail, we also confirmed that our new 
approach properly reproduces the other independent color-ordered 
five-gluon helicity amplitude.

Next we computed the six- and seven-point QCD amplitudes
$A_{6;1}^{\rm QCD}(1^-,2^-, 3^+, 4^+, 5^+, 6^+)$ and $A_{7;1}^{\rm
QCD}(1^-,2^-,3^+, 4^+, 5^+, 6^+,7^+)$.  The rational terms of these
amplitudes had not been computed previously.  Our computations of
these terms use as input lower-point
amplitudes~\cite{BKStringBased,KunsztEtAl,GGGGG,Mahlon,OnShellRecurrenceI},
and the cut-containing terms of the amplitudes under consideration,
obtained previously via the unitarity method~\cite{Neq1Oneloop}. For
the six-point case we presented a compact expression for the complete
amplitude.

Another possible approach to obtaining complete loop amplitudes is via
$D$-dimensional
unitarity~\cite{Massive,SelfDualYM,DDimUnitarity,BMST}.  It would be
worthwhile to corroborate the results of this paper starting from the
known $D$-dimensional tree
amplitudes~\cite{Massive,SelfDualYM,BadgerMassive}.  It would be also
be desirable to develop a first-principles understanding of loop-level
factorization with complex momenta, instead of the heuristic one of
refs.~\cite{OnShellRecurrenceI,Qpap}.

The computation of rational function terms has been a bottleneck for
calculating one-loop amplitudes in non-supersymmetric gauge theories
with six or more external particles.  We expect the technique
discussed in this paper to apply to all one-loop multi-parton amplitudes
in QCD with massless quarks.  It should also work, without modification, 
for amplitudes that contain external massive vector bosons, 
or Higgs bosons (in the limit of a large top-quark mass), 
in addition to massless partons.
Finally, we expect suitable modifications of the method to be 
applicable to processes with massive particles propagating in the loop.

\section*{Acknowledgments}

We thank Iosif Bena, Darren Forde and especially Carola Berger for
helpful discussions.  We also thank Keith Ellis, Walter Giele and
Giulia Zanderighi for pointing out an incorrect sign in an earlier
version, in \eqn{Ff6}, which arose in converting the expression from
ref.~\cite{Neq1Oneloop}. We thank Academic Technology Services at UCLA
for computer support.  We also thank the KITP at Santa Barbara for
providing a stimulating environment at the 2004 Collider Physics
Program, helping to inspire the solution presented in this paper.

%%%%%%%%%%%%%%%%%%%%%%%%%%%%%%%%%%%%%%%%%%%%%%%%%%%%%

\end{document}